\documentclass[MSNbibl,nameyear,dvips]{arxstspdf}
\usepackage{flushend}
\usepackage{stfloats}

\volume{27}
\issue{1}
\pubyear{2012}
\firstpage{31}
\lastpage{50}
\doi{10.1214/11-STS355}

\makeatletter
\renewcommand{\epsilon}{\varepsilon}

\newtheorem{theorem}{Theorem}
\newproclaim{remark}{Remark}

\newcommand{\bet}{\begin{theorem}}
\newcommand{\eet}{\end{theorem}}
\newcommand{\RR}{\mathbb{R}}
\newcommand{\al}{\alpha}
\newcommand{\om}{\omega}
\newcommand{\de}{\delta}
\newcommand{\ep}{\varepsilon}
\newcommand{\lam}{\lambda}
\newcommand{\goto}{\rightarrow}
\newcommand{\hf}{\frac{1}{2}}
\newcommand{\sgn}{\operatorname{sgn}}
\newcommand{\Bb}{B^\al_{p,q}(M)}
\newcommand{\Bbi}{B^{\al_1}_{p_1,q_1}(M_1)}
\newcommand{\Bbii}{B^{\al_2}_{p_2,q_2}(M_2)}

\newcommand{\sd}{\mathcal{D}}
\newcommand{\si}{\mathcal{I}}
\makeatother

\begin{document}
\begin{frontmatter}
\vspace*{6pt}
\title{Minimax and Adaptive Inference in~Nonparametric Function Estimation}
\runtitle{Minimax and Adaptive Inference}

\begin{aug}
\author{\fnms{T. Tony} \snm{Cai}\corref{}\ead[label=e1]{tcai@wharton.upenn.edu}}
\runauthor{T. T. Cai}

\affiliation{University of Pennsylvania}

\address{T. Tony Cai is the Dorothy Silberberg Professor of Statistics, Department of
Statistics, The Wharton School,
University of Pennsylvania,
Philadelphia, Pennsylvania 19104, USA \printead{e1}.}

\end{aug}

%
\begin{abstract}
Since Stein's 1956 seminal paper, shrinkage has played a~fundamental
role in both parametric and nonparametric inference. This article
discusses minimaxity and adaptive minimaxity in nonparametric function
estimation. Three interrelated problems, function estimation under
global integrated squared error, estimation under pointwise squared
error, and nonparametric confidence intervals, are considered.
Shrinkage is pivotal in the development of both the minimax theory and
the adaptation theory.

While the three problems are closely connected and the minimax theories
bear some similarities, the adaptation theories are strikingly
different. For example, in a sharp contrast to adaptive point
estimation, in many
common settings there do not exist nonparametric confidence intervals
that adapt to the unknown smoothness of the underlying function.
A concise account of these theories is given.
The connections as well as differences among these problems are
discussed and illustrated through examples.
\end{abstract}

%

\end{frontmatter}

\section{Introduction}
\label{intro.sec}

The multivariate normal mean model
%
%
\begin{eqnarray}\label{normal.model0} x_i =
\theta_i + \sigma z_i,\quad z_i \stackrel{\mathrm{i.i.d.}}{\sim}
N(0, 1),\nonumber\\ [-8pt]\\ [-8pt]
\eqntext{i=1, \ldots, m,}
\end{eqnarray}
occupies a central position in parametric inference.
In his seminal paper, Stein (\citeyear{Ste56}) showed that, when the dimension
$m\ge3$, the usual maximum likelihood estimator $Y=(y_i)$ of the
normal mean is inadmissible under mean squared error
%
%
\begin{equation}\label{risk}
R(\hat\theta, \theta) = \frac{1}{m} \sum E(\hat\theta_i - \theta_i)^2,
\end{equation}
and demonstrated that significant gain can be
achie\-ved by using shrinkage estimators. Since then shrinkage has become
an indispensable technique in statistical inference, both in parametric
and nonparametric settings.

This article considers minimaxity and adaptive~mi\-nimaxity in
nonparametric function estimation. Spe\-cifically, we discuss three
interrelated problems: function estimation under global integrated
squared error, estimation under pointwise squared error, and
nonparametric confidence intervals. The goal is to~gi\-ve a concise
account of important results in both~the minimax theory and adaptation
theory for each problem. The connections as well as differences among
these problems will be discussed and illustrated\break through examples.
Shrinkage methods, including linear shrinkage, separable rules,
thresholding and bloc\-kwise James--Stein procedures, figure prominently
in the discussion.


A primary focus in nonparametric function estimation is the
construction of adaptive procedures. The goal of adaptive inference is
to construct a single procedure that achieves optimality
simultaneously over a~collection of parameter spaces. Informally an
adaptive procedure automatically adjusts to the smoothness properties
of the underlying function. A common way to evaluate such a procedure
is to compare its maximum risk over each parameter space in the
collection with the corresponding minimax risk.

As a step toward the goal of adaptive inference, one should first focus
attention on the more concrete goal of developing a minimax theory over
a given parameter space. This theory is now well developed particularly
in the white noise with drift model:
%
%
\begin{equation}\label{w.model}
\hspace*{10pt}dY(t) = f(t)\,dt + n^{-1/2}\,dW(t),\quad0 \le t \le1,
\end{equation}
where $W(t)$ is a
standard Brownian motion. This canonical white noise model is
asymptotically equivalent to the conventional nonparametric regression
where one observes $(x_k, y_k)$ with
\[
y_k = f(x_k) + z_k,\quad z_k \stackrel{\mathrm{i.i.d.}}{\sim} N(0,
1),\quad k = 1,
\ldots, n,
\]
where $x_k = k/n$ in the fixed design case and $x_k
\stackrel{\mathrm{i.i.d.}}{\sim} \operatorname{Uniform}(0, 1)$ in
the case of random design.
The parameter $n$ in the white noise model (\ref{w.model}) corresponds
to the
sample size in the regression model. See Brown and Low (\citeyear
{BroLow96N1}) and
Brown et al. (\citeyear{Broetal02}). There is also a
slightly less direct
equivalence to density estimation and spectrum estimation. See Nussbaum
(\citeyear{Nus96}), Klemel\"{a} and Nussbaum (\citeyear{KleNus99})
and Brown et al. (\citeyear{Broetal04}).

Let $\{\beta_i(t), i\in\mathcal{I}\}$ be an orthonormal basis\break of
$L^2[0, 1]$ and let $y_i=\int\beta_i (t)\,dY_n(t)$ and $\theta
_i=\break\int
f(t) \beta_i(t)\,dt$. Then the white noise model (\ref{w.model}) is
equivalent to the following infinite-dimensional Gaussian sequence
model
%
%
\begin{equation}\label{seq.model0}
\hspace*{10pt}y_i = \theta_i+ n^{-1/2} z_i,\quad z_i \stackrel
{\mathrm{i.i.d.}}{\sim}
N(0, 1),\quad i\in\mathcal{I}.
\end{equation}
An estimator
$\hat
\theta$ of the mean sequence $\theta$ directly provides an estimator
$\hat f (t) = \sum_{i\in\mathcal{I}} \hat\theta_i \beta_i(t)$ of the
function $f$ in the white noise model and vice versa. Hence, the
function estimation model is closely related to the classical
multivariate normal mean mo\-del~(\ref{normal.model0}). In these
infinite-dimensional problems it is necessary to restrict the parameter
set to be a compact subset of $\ell^2$, the space of square summable
sequences (or a compact subset of $L^2$, the space of square integrable
functions, in the case of the white noise model). In contrast, the
parameter set in the finite dimensional problem is typically all of
$\RR^m$.

Two of the most common ways of evaluating the performance of
nonparametric function estimators are integrated squared error and
pointwise squared error. Integrated squared error is used as a global
measure of accuracy whereas pointwise squared error gives a local
measure of loss. Minimax theory for both of these cases has been
developed. We shall begin our discussion on minimax theory for
estimation under integrated squared error. What follows will be
elaborated in Section \ref{minimax.sec}. Pinsker (\citeyear{Pin})
made a major
breakthrough in nonparametric function estimation theory by giving a
complete and explicit solution to the problem of minimax estimation
over an ellipsoid under integrated squared error loss. Pinsker derived
the minimax linear estimator and showed that the minimax risk is equal
to the linear minimax risk asymptotically. Together these results yield
the first precise evaluation of the asymptotic minimax risk in
nonparametric function estimation. Donoho, Liu and McGibbon (\citeyear
{DonLiuMac90})
considered certain more general quadratically convex parameter spaces
and showed that the linear minimax risk is within a small constant of
the minimax risk. Furthermore, they also showed the limitations of
linear procedures when the parameter space is not quadratically convex.
Donoho and Johnstone (\citeyear{DonJoh98}) studied minimax estimation
over Besov balls
which include cases that are not quadratically convex. Besov spaces are
a~very rich class of function spaces that are commonly used to model
functions of inhomogeneous smoothness in functional analysis,
statistics and signal processing. They also contain as special cases
many traditional smoothness spaces such as H\"{o}lder and Sobolev
spaces. The results of Donoho and Johnstone marked another major
advance in the minimax estimation theory. In this setting it is shown
that nonlinearity is essential for achieving minimaxity or even the
minimax rate.
Moreover, it is shown that the risk of the optimal coordinatewise
thresholding rule is within a constant factor of the minimax risk.

The problem of estimating a function under pointwise squared error will
be discussed in Section \ref{pointwise.sec}. This problem can be
considered as a special case of estimating a linear functional. The
minimax theory for estimating a linear functional\vadjust{\goodbreak} over a convex
parameter space has been well developed in Ibragimov and Hasminskii
(\citeyear{IbrHas84}), Donoho and Liu (\citeyear{DonLiu91}) and
Donoho (\citeyear{Don94}). In particular, the
minimax difficulty of estimation is captured by a geometric quantity,
the modulus of continuity, and the optimal linear shrinkage estimator
is within a $1.25$ multiple of the minimax risk. Cai and Low (\citeyear
{CaiLow04N1})
extended this minimax theory to nonconvex parameter spaces. In this
case, although the minimax rate of convergence is still determined by
the modulus of continuity, optimal linear procedures can be arbitrarily
far from being minimax and nonlinearity is necessary for minimax
estimation.

The theory of adaptive estimation depends strong\-ly on how risk is
measured. When the performance is measured globally sharp adaptation
can often be achieved. That is, one can attain the minimax risk
over a collection of parameter spaces simultaneously.
In particular, Efromovich and Pinsker (\citeyear{EfrPin84})
constructed sharp adaptive
estimators over a range of Sobolev spaces. Recent results on rate
adaptive estimators focus on the more general Besov spaces. See, for
example, Donoho and Johnstone (\citeyear{DonJoh95}), Cai (\citeyear
{Cai99}), Johnstone and
Silverman (\citeyear{JohSil05}) and Zhang (\citeyear{Zha05}). In
particular, Zhang (\citeyear{Zha05})
developed general empirical Bayes methods which are asymptotically
sharp minimax simultaneously over a wide collection of Besov balls.
Adaptive estimation under the global loss will be discussed in Section
\ref{adaptive.sec}. While separable rules are optimal for minimax
estimation, they cannot be rate adaptive. Information pooling is a
necessity for achieving adaptivity. Block thresholding provides a
convenient and effective tool for information pooling. We discuss in
detail block thresholding rules via the approach of ideal adaptation
with an oracle. Through block thresholding, many shrinkage estimators
developed in the normal decision theory can be used for nonparametric
function estimation. In this sense block thresholding serves as a
bridge between the classical theory and the modern function estimation
theory.

Under pointwise risk it is often the case that sharp adaptation is not
possible and a penalty, usually a~logarithmic factor, must be paid for
not knowing the smoothness. Important work in this area began with
Lepski (\citeyear{Lep90}) where attention focused on a collection of Lipschitz
classes. Brown and Low (\citeyear{BroLow96N2}) obtained similar
results using a
constrained risk inequality, Tsybakov (\citeyear{Tsy98}) investigated pointwise
adaptation over Sobolev classes and Cai (\citeyear{Cai03}) considered Besov
spaces. Efromovich and Low (\citeyear{EfrLow94}) studied\vadjust{\goodbreak} estimation of linear
functionals over a nested sequence of symmetric sets. A general
adaptation theory for estimating linear functionals is given in Cai and
Low (\citeyear{CaiLow05N1}). This theory gives a geometric
characterization of the
adaptation problem analogous to that given by Donoho (\citeyear
{Don94}) for minimax
theory. The adaptation theory describes exactly when rate adaptive
estimators exist and when they do not exist the theory provides a
general construction of estimators with the minimum adaptation cost.

In addition to point estimation, confidence sets also play a
fundamental role in statistical inference. The construction of
nonparametric confidence sets is an important and challenging problem.
In Section~\ref{CI.sec} we consider nonparametric confidence sets with
a~particular focus on confidence intervals. Other confidence sets such
as confidence balls and confidence bands have also been discussed in
the literature. A minimax theory of confidence intervals for linear
functionals was given in Donoho (\citeyear{Don94}) when the parameter
space is
assumed to be convex. Donoho (\citeyear{Don94}) constructed fixed
length intervals
centered at linear estimators which have length within a small constant
factor of the minimax expected length. Cai and Low (\citeyear
{CaiLow04N2}) extended the
minimax theory for parameter spaces that are finite unions of convex
sets. In this case it is shown that optimal confidence intervals
centered at linear estimators can have expected length much larger
than the minimax expected length. It is thus essential to center the
confidence interval at a nonlinear estimator in order to achieve
minimaxity over nonconvex parameter spaces.

An adaptation theory for confidence intervals was developed in Cai and
Low (\citeyear{CaiLow04N1}). When attention is focused on adaptive
inference there are
some striking differences between adaptive estimation and adaptive
confidence intervals. As mentioned earlier, sharp adaptation is often
possible under integrated squared error and the cost of adaptation is
typically a logarithmic factor under pointwise squared error. In
contrast, in many common cases the cost of adaptation for confidence
intervals is so high that adaptation becomes essentially impossible.

There is also a conspicuous difference between confidence intervals in
parametric and nonparametric settings. To construct a confidence
interval in parametric inference, a virtually universal technique is to
first derive an optimal estimator of a parameter and then construct a
confidence interval centered at this optimal\vadjust{\goodbreak} estimator. It is often the
case that such a method leads to an optimal confidence interval for the
parameter. This is also a common practice in nonparametric function
estimation. However, somewhat surprisingly, centering confidence
intervals at optimally adaptive estimators in general yield suboptimal
confidence procedures (Cai and Low, \citeyear{CaiLow05N3}): Either the
resulting
interval has poor coverage probability or it is unnecessarily long.

The paper is organized as follows. We begin with minimax estimation
under global integrated squared error loss. Section \ref{minimax.sec}
focuses on the important results developed in Pinsker (\citeyear
{Pin}), Donoho,
Liu and McGibbon (\citeyear{DonLiuMac90}) and Donoho and Johnstone
(\citeyear{DonJoh98}) on linear
minimaxity, separable rules and minimaxity. Section \ref{adaptive.sec}
considers adaptive estimation under the global loss. The performance of
separable rules~is studied in the context of adaptive estimation. The~re\-sults show
that separable rules cannot be rate adaptive and
information pooling is essential for adaptive estimation. We then
discuss block thresholding rules using an oracle approach. Section
\ref{pointwise.sec} considers minimax and adaptive estimation under
pointwise squared error loss and the construction of minimax and
adaptive confidence intervals is treated in Section \ref{CI.sec}. The
paper is concluded with discussions in Section \ref{discussion.sec}.

\section{Linear Minimaxity, Separable Rules and~Minimaxity}
\label{minimax.sec}

Minimax theory has been well developed in the Gaussian sequence model
(and, equivalently, the whi\-te noise model). Two classes of estimators,
namely, linear shrinkage rules and separable rules, figure pro\-minently
in the development of the theory. In this section we consider minimax
estimation under global mean integrated squared error (MISE)
%
%
\begin{eqnarray}\label{g.risk}
R(\hat f, f) &=& E_f\|\hat f - f\|_2^2\nonumber\\ [-8pt]\\ [-8pt]
& =& E_f\int_0^1\bigl(\hat f(t)
-f(t)\bigr)^2\,dt\nonumber
\end{eqnarray}
for the function estimation model
(\ref{w.model}) and
\[
R(\hat\theta, \theta) = E_\theta\| \hat\theta-\theta\|^2_2
\]
for the sequence estimation model (\ref{seq.model0}). Because of the
isometry of the risks $R(\hat f, f) = R(\hat\theta, \theta)$ we
shall focus on the sequence model (\ref{seq.model0}) in this section.
The performance of an estimator $\hat\theta$ over a parame\-ter set~$\mathcal{F}$ is measured by its maximum risk
\[
R_n(\hat\theta, \mathcal{F}) = \sup_{\theta\in\mathcal{F}}
E_\theta
\|\hat\theta- \theta\|_2^2\vadjust{\goodbreak}
\]
and the benchmark is the minimax risk
\[
R_n^*(\mathcal{F}) = \inf_{\hat\theta}\sup_{\theta\in\mathcal{F}}
E_\theta
\|\hat\theta- \theta\|_2^2.
\]
When attention is restricted to linear
procedures, we consider the linear minimax risk
\[
R^L_n({\mathcal{F}}) = \inf_{\hat\theta\ \mathrm{linear}} \sup
_{\theta\in
\mathcal{F}}E_\theta\|\hat\theta- \theta\|^2_2.
\]

In this section we give a concise account of some of the most important
results in the minimax estimation theory without getting into too much
technical detail. We refer interested readers to Iain Johnstone's
monograph (Johnstone, \citeyear{Joh}) for a detailed discussion on
these and
other related results.

\subsection{Linear Minimaxity}
\label{linear.sec}

Linear estimators and linear minimax risk occupy a special place in the
development of nonparametric function estimation theory. Linear
procedures are appealing because of their simplicity and linear minimax
risk is easier to evaluate than the minimax risk. For example, for
linear estimation over solid and orthosymmetric parameter spaces it
suffices to focus on simple diagonal linear estimators of the form
$\hat\theta_i = w_i y_i$ where $w_i$ is a constant. Furthermore, in
many settings the optimal linear procedure is asymptotically minimax or
within a small constant of the minimax risk. See, for example, Pinsker
(\citeyear{Pin}) and Donoho, Liu and McGibbon (\citeyear{DonLiuMac90}).
In this section we shall follow the historical development of the
linear minimax theory by discussing the theory in the order of
ellipsoids, quadratically convex classes and Besov classes.

\subsubsection*{Linear minimaxity over ellipsoids}

Pinsker (\citeyear{Pin}) considered minimax estimation over an ellipsoid
%
%
\begin{equation}
\mathcal{F}
= \Biggl\{\theta\dvtx \sum_{i=1}^\infty a_i^2 \theta_i^2 \le
M\Biggr\},
\end{equation}
where
$a_i \ge0$ and $a_i \goto\infty$. Since the ellipsoid $\mathcal{F}$ is
symmetric, the linear minimax risk is attained by the optimal diagonal
linear estimator of the form $\hat\theta(w) = (w_i y_i)$ where
$w=(w_i)\in\ell^2$ with $0\le w_i\le1$ is a sequence of weights. That
is,
%
%
\begin{equation}\label{pinsker1}
R^L_n({\mathcal{F}}) = \inf_w \sup_{\theta\in
\mathcal{F}}
E_\theta\|\hat\theta(w) -\theta\|_2^2 .
\end{equation}
The RHS of
(\ref{pinsker1}) is easy to evaluate. Note that
\[
E_\theta\|\hat\theta(w) -\theta\|_2^2 = \sum_{i=1}^\infty
\bigl(n^{-1}w_i^2 + (1-w_i)^2 \theta_i^2\bigr).
\]
Hence, the linear minimax risk
%
%
\begin{eqnarray}\label{max.linear.risk}
\hspace*{10pt}R^L_n({\mathcal{F}}) &=& \inf_w \sup_{\theta\in
\mathcal{F}} \sum
_{i=1}^\infty
\bigl(n^{-1}w_i^2 + (1-w_i)^2 \theta_i^2\bigr)\nonumber\\ [-8pt]\\ [-8pt]
\hspace*{10pt}&=& \sup_{\theta\in\mathcal{F}}
\sum_{i=1}^\infty{n^{-1}\theta_i^2\over n^{-1} + \theta
_i^2}.\nonumber
\end{eqnarray}
For
any real number $x$, write $(x)_+$ for $\max(x, 0)$. The Lagrange
multiplier method shows that the maximum on the RHS of
(\ref{max.linear.risk}) is attained at $\theta_i^2 = n^{-1}(\mu/\break a_i -
1)_+$, where the parameter $\mu$ is determined by the constraint
$\sum_{i=1}^\infty a_i^2 \theta_i^2 = M$, which is equivalent
to\looseness=-1
\[
n^{-1} \sum_{i=1}^\infty a_i (\mu- a_i)_+ = M.
\]\looseness=0
The minimax linear estimator is given by $\hat\theta_{\mathrm
{l.minimax}} =
(\hat\theta_i)$ with
%
%
\begin{equation}\label{l.minimax.est} \hat\theta_i =
(1-a_i/\mu)_+ y_i
\end{equation}
and the linear minimax risk is
%
%
\begin{equation}\label{pinsker2}
R^L_n({\mathcal{F}}) = n^{-1} \sum_{i=1}^\infty
(1-a_i/\mu)_+.
\end{equation}
A remarkable result of Pinsker (\citeyear{Pin}) is that for
ellipsoidal $\mathcal{F}$ the linear minimax risk is asymptotically
equal to
the minimax risk, that is,
\[
R_n^*(\mathcal{F}) = R_n^L(\mathcal{F})\bigl(1+o(1)\bigr).
\]
Therefore, the minimax linear estimator $\hat\theta_{\mathrm{l.minimax}}$
given in (\ref{l.minimax.est}) is asymptotically minimax and the
minimax risk is equal to the RHS of (\ref{pinsker2}) asymptotically.

In the case of special interest where the parameter space is a Sobolev
ball
\[
\Theta^\al_2(M) = \Biggl\{\theta\dvtx \sum_{k=1}^\infty(2\pi
k)^{2\al}
(\theta_{2k}^2 + \theta_{2k+1}^2) \le M\Biggr\}
\]
(which corresponds to a Sobolev ball in the function space under the
usual trigonometric basis), the asymptotic minimax risk and the linear
minimax risk can be evaluated explicitly as
%
%
\begin{eqnarray}\label{pinsker3}
R_n^*(\Theta^\al_2(M))&=&
R_n^L(\Theta^\al_2(M))\bigl(1+o(1)\bigr)\nonumber\\
&=&\pi^{-2\al/(1+2\al)} M^{2/(1+2\al)} P_\al\\
&&{}\cdot
n^{-2\al/(1+2\al)}
\bigl(1+o(1)\bigr),\nonumber
\end{eqnarray}
where
\[
P_\al= \biggl(\frac{\al}{1+\al}\biggr)^{2\al/(1+2\al)} (1+2\al
)^{1/(1+2\al)}
\]
is the Pinsker constant. This is the first exact evaluation of the
asymptotic minimax\vadjust{\goodbreak} risk in the nonparametric function estimation
problem. See also Efromovich and Pinsker (\citeyear{EfrPin82}) and
Nussbaum (\citeyear{Nus85}).

Pinsker's results represent a major contribution to nonparametric
function estimation theory. Together they offer a complete and explicit
solution to the problem of minimax estimation over ellipsoids.

\subsubsection*{Linear minimaxity over quadratically convex clas\-ses}

Donoho, Liu and MacGibbon (\citeyear{DonLiuMac90}) considered certain
more general
quadratically convex parameter spaces. To discuss their results in more
detail, we need first to introduce some terminology.

A parameter space $\mathcal{F}$ is called solid and orthosymmetric if
$\theta=
(\theta_1, \ldots, \theta_k, \ldots)\in\mathcal{F}$ implies that
$\xi\in
\mathcal{F}$
if $|\xi_i|\le|\theta_i|$ for all $i$. A set $\mathcal{F}$ is called
quadratically convex if the set $\{(\theta_i^2)_{i=1}^\infty\dvtx
\theta
\in\mathcal{F}\}$ is convex. The quadratic convex hull of a set
$\mathcal{F}$ is
defined as
%
%
\begin{equation}
\hspace*{20pt}\operatorname{Q.Hull}(\mathcal{F}) = \{(\theta
_i)_{i=1}^\infty\dvtx
(\theta_i^2)_{i=1}^\infty\in\operatorname{Hull}(\mathcal{F}_+^2)\},
\end{equation}
where
$\mathcal{F}_+^2 =
\{(\theta_i^2)_{i=1}^\infty\dvtx(\theta_i)_{i=1}^\infty\in
\mathcal{F},
\theta_i
\ge0\ \forall i\}$ and\break $\operatorname{Hull}(\mathcal{F}_+^2)$ denotes the
closed convex
hull of the set~$\mathcal{F}_+^2$.\vspace*{1pt}

Donoho, Liu and MacGibbon (\citeyear{DonLiuMac90}) showed that for all solid
orthosymmetric, compact and quadratically convex parameter spaces
$\mathcal{F}$
the linear minimax risk is within a $1.25$ factor of the minimax risk,
that is,
%
%
\begin{equation}\label{1.25}
R_n^L(\mathcal{F}) \le1.25
R_n^*(\mathcal{F}).
\end{equation}
Hence,
the optimal linear procedure cannot be substantially improved by a
nonlinear estimator. Donoho, Liu and MacGibbon (\citeyear{DonLiuMac90}) proceeded by
first solving an infinite-dimensional hyperrectangle problem where the
parameter space $\mathcal{F}$ is of the form
%
%
\begin{equation}\label{hyper}
\mathcal{F}= \{
\theta\dvtx|\theta_i| \le\tau_i, i=1, 2, \ldots\}
\end{equation}
with $\sum_i \tau_i^2
<\infty$. The traditional H\"{o}lder smoothness constraint in the
function space corresponds to a hyperrectangle constraint in the
sequence space with a suitably chosen $(\tau_i)$. See, for example,
Meyer (\citeyear{Mey92}). The problem of estimation over a
hyperrectangle is solved
by reducing it to coordinatewise one-dimensional bounded normal mean
problems.

Consider estimating a bounded normal mean $\theta\in\RR$ based on one
observation $y \sim N(\theta, \sigma^2)$ with the prior knowledge that
$|\theta|\le\tau$. It is easy to show that the minimax linear
estimator of the bounded normal mean $\theta$ is
\[
\delta^L(y) = \frac{\tau^2}{\tau^2+\sigma^2} y
\]
and the minimax linear risk is
\[
\rho^L(\tau, \sigma) \equiv\inf_{\delta\ \mathrm{linear}}
\sup_{|\theta|\le
\tau}E_\theta\bigl(\delta(y) - \theta\bigr)^2 = \frac{\tau
^2\sigma^2}{\tau^2 +\sigma^2}.\vadjust{\goodbreak}
\]
Denote the minimax risk for estimating the bounded normal mean $\theta$
by $\rho^*(\tau, \sigma)$. Let $\mu^*$ be the maximum value of the
ratio of $\rho^L(\tau, \sigma)$ and $\rho^*(\tau, \sigma)$, that is,
%
%
\begin{equation}\label{IH.constant}
\mu^* = \sup_{\tau, \sigma}
\frac{\rho^L(\tau,\sigma)}{\rho^*(\tau,\sigma)}.
\end{equation}
The constant $\mu^*$ is called
the Ibragimov--Hasminskii constant. Ibragimov and Hasminskii (\citeyear
{IbrHas84})
studied the properties of the ratio $\rho^L(\tau,
\sigma)/\rho^*(\tau,\sigma)$ and showed that the constant $\mu^*$ is
finite. Donoho, Liu and MacGibbon (\citeyear{DonLiuMac90}) proved that
$\mu^*$ is in fact
less than or equal to $1.25$.

For estimation of $\theta$ over the hyperrectangle $\mathcal{F}$
given in
(\ref{hyper}) based on the sequence model (\ref{seq.model0}), due to
the independence of the observations $y_i$ and the independent
constraints on $\theta_i$, it is not difficult to see that the minimax
problem is separable. That is, the
minimax (linear) estimator can be obtained through coordinatewise
minimax (linear) estimation. Hence,
\begin{eqnarray*}
R^L_n(\mathcal{F}) &=& \sum_{i=1}^\infty\rho^L(\tau_i, n^{-1})
\quad\mbox{and}\\ R^*_n(\mathcal{F})
& =& \sum_{i=1}^\infty\rho^*(\tau
_i, n^{-1})
\end{eqnarray*}
and, consequently, for hyperrectangle $\mathcal{F}$,
%
%
\begin{equation}\label{hyper1.25}
R^L_n(\mathcal{F}) \le\mu^* R^*_n(\mathcal{F}) \le1.25
R^*_n(\mathcal{F}).
\end{equation}
A key
step in
solving the more general quadratically convex problem is to show that
the difficulty for the linear estimators over the quadratically convex
parameter space is in fact equal to the difficulty for the linear
estimators of the hardest rectangular subproblem. Then (\ref{1.25})
follows directly from (\ref{hyper1.25}).

In addition, Donoho, Liu and MacGibbon (\citeyear{DonLiuMac90}) also
showed that the
linear minimax risk over a solid compact orthosymmetric set $\mathcal
{F}$ is
equal to that over the quadratic convex hull of $\mathcal{F}$,
%
%
\begin{equation}\label{NQ.risk}
R^L_n(\mathcal{F}) = R^L_n(\operatorname{Q.Hull}(\mathcal{F})).
\end{equation}
This result
indicates that although the optimal linear estimator is near minimax
over quadratically convex parameter spaces, linear procedures have
serious limitations when the parameter space $\mathcal{F}$ is not quadratically
convex, especially when the quadratic convex hull of $\mathcal{F}$ is much
larger than $\mathcal{F}$ itself. Such is the case in wavelet function
estimation over certain Besov balls and in estimation of a sparse
normal mean.

\subsubsection*{Linear minimaxity over Besov classes}

We now\break turn to wavelet estimation over Besov balls. It is more
convenient to use double indices and write the sequence model
(\ref{seq.model0}) as
%
%
\begin{eqnarray}\label{seq.model}
y_{j,k} &=& \theta_{j,k}+ n^{-1/2}
z_{j,k},\nonumber\\ [-8pt]\\ [-8pt]
 z_{j,k} &\stackrel{\mathrm{i.i.d.}}{\sim}& N(0, 1),\quad
(j,k)\in\si,\nonumber
\end{eqnarray}
where the index set $\si= \{(j, k)\dvtx k = 1,
\ldots, 2^j, j = 0,\break 1, \ldots\}$.
The Besov seminorm $\|\cdot\|_{b^\al_{p,q}}$ in the sequence space is
then defined as
%
%
\begin{equation}\label{Besov.norm}
\hspace*{25pt}\|\theta\|_{b_{p,q}^\alpha} =
\Biggl(\sum_{j=0}^\infty\Biggl(2^{js} \Biggl(\sum
_{k=1}^{2^j}|\theta_{j,k}|^p
\Biggr)^{1/p}\Biggr)^q\Biggr)^{1/q},
\end{equation}
where $s = \alpha+ {1
\over2} - {1 \over p}$. We shall assume throughout the paper that $p,
q, \al, s > 0$. The Besov ball $\Bb$ is defined as a ball of
radius $M$ under this seminorm, that is,
%
%
\begin{equation}\label{Besov.ball}
B_{p,q}^\alpha(M) = \{ \theta\dvtx\|\theta\|_{b_{p,q}^\alpha}
\le M\}.
\end{equation}
Besov spaces are a very rich class of function spaces
and occur naturally in many areas of analysis. Besov spaces contain as
special cases several traditional smoothness spaces such as H\"{o}lder
and Sobolev spa\-ces. For example, a H\"{o}lder space is a Besov space
with $p=q = \infty$ and a Sobolev space is a Besov space with $p=q=2$.
Full details of Besov spaces are given, for example, in Triebel
(\citeyear{Tri92})
and DeVore and Lorentz (\citeyear{DeVLor93}). See Meyer (\citeyear
{Mey92}) and Daube\-chies (\citeyear{Dau92})
for wavelets and correspondence between function spaces and sequence
spaces.

It is easy to verify that for $p \ge2$ the Besov ball $\Bb$ is
quadratically convex and when $p < 2$,
%
%
\begin{equation}\label{Bb.rate}
\operatorname{Q.Hull}(\Bb) = B_{2,q}^s(M),
\end{equation}
where again $s={\alpha+ \hf- {1 \over p}}$. Besov spaces with $p < 2$
contain functions of a high degree of spatial inhomogeneity. See, for
example, Triebel (\citeyear{Tri92}), Meyer (\citeyear{Mey92}) and
DeVore and Lorentz (\citeyear{DeVLor93}).
Equa-\break tions~(\ref{Bb.rate}) and (\ref{NQ.risk}) together imply that for
the Besov ball $\Bb$ with $p<2$,
%
%
\begin{eqnarray}\label{Besov.linear.risk0}
R^L_n(\Bb) &=& R^L_n(\operatorname{Q.Hull}(\Bb)) \nonumber\\
[-8pt]\\ [-8pt]
&=& R^L_n(B_{2,q}^s(M)).\nonumber
\end{eqnarray}
In
particular, for $p < 2$ the linear minimax risk over $\Bb$ converges at
the same rate as the minimax risk over $B_{2,q}^s(M)$. As we will see
in Section \ref{separable.minimaxity.sec}, the minimax risk over $\Bb$
converges at the rate of $n^{-2\al/(1+2\al)}$ (Donoho and
Johnstone, \citeyear{DonJoh98}). Since\vspace*{1pt} $s < \al$ for $p
< 2$, $n^{-2s/(1 + 2s)}
\gg n^{-2\al/(1+2\al)}$ and so the linear minimax risk over a
Besov ball $B_{p,q}^{\al}(M)$ with $p<2$ is substantially larger than
the minimax risk. Therefore, the optimal linear estimator can be
significantly outperformed by a nonlinear procedure. Intuitively,
linear estimators do not perform well when the underlying functions are
spatially inhomogeneous. In this case it is thus no longer desirable to
restrict attention to the class of linear estimators.\looseness=1

\begin{remark}
It is interesting to note that a~\mbox{similar} phenomenon also arises in
the estimation of a~quadra\-tic functional. Cai and Low (\citeyear
{CaiLow05N2}) showed
that for estimating the quadratic functional $Q(\theta) =
\sum_{i=1}^\infty\theta_i^2$ in the sequence model (\ref{seq.model0}),
the minimax quadratic risk over a solid orthosymmetric parameter space
$\mathcal{F}$ equals the minimax quadratic risk over the quadratic
convex hull
of $\mathcal{F}$. Consequently, the optimal quadratic estimator of the
quadratic functional $Q(\theta)$ is far from being minimax over a Besov
ball $\Bb$ with $p<2$.
\end{remark}

\subsection{Separable Rules and Minimaxity}
\label{separable.minimaxity.sec}

The shortcoming of linear procedures shows that nonlinearity is a
necessity for achieving minimaxity over parameter spaces that are not
quadratically convex, such as Besov balls $\Bb$ with $p<2$. Separable
rules, which apply nonlinearity to individual coordinates separately,
are a natural generalization of the linear shrinkage rules.
Separable rules play a fundamental role in minimax estimation over
parameter spaces that are not quadratically convex in a way similar to
the role played by the linear estimators over the more conventional
parametric spaces such as ellipsoids and hyperrectangles.

Under the sequence model (\ref{seq.model}), an estimator $\de=
(\de_{j,k})$ is {\it separable} if for all $(j, k)\in\si$,
$\de_{j,k}$ depends solely on $y_{j,k}$, not on any other $y$'s. We
shall denote by $\mathcal{S}$ the collection of all separable rules. Well-known
examples of separable rules include the traditional diagonal linear
estimators, term-by-term thresholding estimators and Bayes estimators
derived from independent priors. Separable rules are attractive because
of their simplicity and intuitive appeal. More importantly, separable
rules are minimax for a wide range of parameter spaces. In an important
paper, Donoho and Johnstone (\citeyear{DonJoh98}) pioneered the study
of separable
rules in minimax estimation over the Besov ball $\Bb$ under the
sequence model (\ref{seq.model}). Zhang\vadjust{\goodbreak} (\citeyear{Zha05}) further
studied the
class of separable rules in the context of sharp adaptation over the
full scale of Besov balls using general empirical Bayes methods.

Donoho and Johnstone (\citeyear{DonJoh98}) began by first solving the following
minimax Bayes estimation problem. Suppose we observe $y=(y_{j,k})$ as
in (\ref{seq.model}) with $\theta=(\theta_{j,k})$ itself a random
vector satisfying a mean constraint
\[
\|\tau\|_{b^\al_{p,q}} \le M,
\]
where
\[
\tau_{j,k} = (E|\theta_{j,k}|^{p\wedge q})^{1/(p\wedge q)},\quad
(j,k)\in\si,
\]
with $p \wedge q = \min(p, q)$. In other words, the
``hard''~con\-straint $\theta\in\Bb$ in the original minimax problem is
replaced by the ``in mean'' constraint $\tau\in\Bb$ in the minimax
Bayes problem. The minimax Bayes risk is defined as
\[
R_n^B (\Bb) = \inf_{\hat\theta} \sup_{\tau\in\Bb} E\|\hat
\theta-
\theta\|_2^2.
\]
Donoho and Johnstone (\citeyear{DonJoh98}) showed that the minimax
Bayes risk
$R_n^B(\Bb)$ is attained by a separable rule $\hat\theta^*=(\hat
\theta^*_{j,k})$ of the form
\[
\hat\theta^*_{j,k} = \delta^*_j(y_{j,k}),
\]
where $\delta^*_j(y_{j,k})$ is a scalar nonlinear\vspace*{-2pt} function of
$y_{j,k}$. Furthermore, when $\al+ \hf> 1/(2\wedge p\wedge q)$, the
mini-max Bayes risk is given by
%
%
\begin{eqnarray}\label{minimax.bayes}
&&R_n^B (\Bb)\nonumber\\
&&\quad=\gamma(M n^{1/2}) M^{2/(1+2\al)} n^{-2\al/(1 +
2\al)}\\
&&\qquad{}\cdot\bigl(1+o(1)\bigr),\quad n \goto\infty,\nonumber
\end{eqnarray}
where $\gamma(\cdot)$ is a continuous, positive, periodic function
of $\log_2(M n^{1/2})$. Moreover, when $p > q$, the minimax risk is
asymptotically equal to the minimax Bayes risk,
\[
R_n^* (\Bb) = R_n^B (\Bb)\bigl(1+o(1)\bigr),
\]
and thus separable rules are minimax. Zhang (\citeyear{Zha05}) further
showed that
the optimal separable rule is asymptotically minimax for general $(p,
q)$.
In particular, these results showed that the minimax rate of
convergence is $n^{-r_*}$ where
%
%
\begin{equation}\label{besov.rate.r} r_* =
\frac{\al}{\al+ 1/2}.
\end{equation}
That is,
\begin{eqnarray*}
0&<& \mathop{\underline{\lim}}_{n \goto\infty} n^{r_*} R_n^* (\Bb)\\
& \le&\mathop{\overline{\lim}}_{n
\goto\infty} n^{r_*} R_n^* (\Bb) < \infty.
\end{eqnarray*}
The linear minimax rate of convergence now follows immediately from
(\ref{1.25}), (\ref{NQ.risk}), (\ref{Bb.rate}) and
(\ref{besov.rate.r}). The linear minimax risk converges at the rate
$n^{-r_\ell}$ where~$r_\ell$ is given by
\begin{eqnarray}
r_\ell= \frac{\al+ (1/p_- - 1/p)}{ \al+ 1/2 +
(1/p_- - 1/p)},\nonumber\\
\eqntext{\mbox{where } p_- = \max(p,2).}
\end{eqnarray}
It is
clear that $r_\ell= r_*$ when $p\ge2$ and $r_\ell< r_*$ when $p <
2$. Hence, nonlinear separable rules can outperform linear estimators
at the level of convergence rates when $p<2$.

\subsection{Rate-Optimal Coordinatewise Thresholding~Estimator}
\label{threshold.sec}

The separable minimax estimator that attains the minimax Bayes risk
(\ref{minimax.bayes}) is not available in closed form. Donoho and
Johnstone (\citeyear{DonJoh98}) showed that attention can be further
restricted to a
simpler coordinatewise thresholding estimator. It is shown that the
optimal term-by-term thresholding estimator is within a~small constant
factor of the minimax risk. It was noted in Donoho and Johnstone
(\citeyear{DonJoh98}) that the constant factor is $\Lambda(p\wedge q)
\le1.6$ for
$p\wedge q = 1$ using computational experiments and $\Lambda(p\wedge q)
\le2.2$ for $p\wedge q = 1$ for the essentially quadratically convex
(and thus less important) case of $p\ge2$. However, no specific rate
optimal thresholding estimator is given in their paper.

We now present a rate-optimal coordinatewise\break thresholding estimator.
Consider the sequence\break model~(\ref{seq.model}). Let $J_0$ and $J$ be
integers satisfying,\vspace*{2pt} \mbox{respectively}, $M^{2/(1+2\al)} n^{1/(1+2\al)}
\leq2^{J_0} <\break 2 M^{2/(1+2\al)} n^{1/(1+2\al)}$ and $n \leq2^J <
2n$. For \mbox{$j \geq J_0+1$}, let
%
%
\begin{equation}\label{thresh}
\lam_j = \sqrt{2 n^{-1} \log(2^{j - J_0})}
\end{equation}
and
let $\eta_\lam(y) = \sgn(y) (|y| -\lam)_+$ be the soft threshold
function. We define the following thresholding estimator:
%
%
\begin{equation}\label{term.est}
\hat\theta_{j,k} = \cases{
y_{j,k}, & if $1 \leq j < J_0$,\cr
\eta_{\lam_j} (y_{j,k}), & if $J_0 \leq j < J$,\cr
0, & if $j \geq J$.
}
\end{equation}
The estimator given in
(\ref{term.est}) is similar to the wavelet estimator given in Delyon
and Juditsky (\citeyear{DelJud96}) for density estimation and
nonparametric regression
over $\Bb$ under the Sobolev norm loss. It differs from the estimator
in Delyon and Juditsky (\citeyear{DelJud96}) in the choice of the
lower and upper
resolution levels $J_0$ and $J$ as well as in the choice of the
thresholds $\lam_j$. The following theorem can be shown using the same
proof as given in Delyon and Juditsky (\citeyear{DelJud96}).

\bet\label{attainment.thm}
The separable estimator $\hat\theta$ given in~(\ref{term.est}) is within a constant factor of the minimax risk over
the Besov ball $\Bb$. That is,
\[
R_n(\hat\theta, \Bb) \le C(\alpha, p, q) R_n^*(\Bb),
\]
where the constant $C(\alpha, p, q)$ depends only on $\alpha$, $p$ and
$q$. In particular, the estimator is minimax rate-optimal,
%
%
\begin{equation}
\hspace*{20pt}\mathop{\overline{\lim}}_{n\goto\infty} n^{2\al/(1+2\al)}
\sup_{\theta\in\Bb}
E\|\hat\theta- \theta\|_2^2 < \infty.
\end{equation}
\eet

\section{Adaptive Estimation through Information Pooling}
\label{adaptive.sec}

Minimax risk provides a useful uniform benchmark for the comparison of
estimators. However, the minimax estimators discussed in Section
\ref{minimax.sec} require some explicit knowledge of the parameter
space which is
unknown in practice. A minimax estimator designed for a specific
parameter space typically performs poorly over another parameter space.
Recent work on nonparametric function estimation has focused attention
on adaptive estimation, with the goal of constructing a single
procedure which is near minimax simultaneously over a collection of
parameter spaces. As mentioned in the \hyperref[intro.sec]{Introduction}, whe\-ther this goal
can be accomplished depends strongly on how risk is measured. When the
performance is measured by the global MISE risk sharp adaptation over
Besov balls can be achieved. In fact, a large number of adaptive
procedures have been developed in the literature. In this section we
consider adaptive estimation under the MISE risk. For reasons of space,
we do not give a comprehensive review of these adaptive estimators. We
shall focus the discussion only on block thresholding which naturally
connects shrinkage rules developed in the classical normal decision
theory with nonparametric function estimation.

Because of the optimal performance of the separable rules in the
minimax estimation setting, we begin in Section \ref{separable.sec} by
studying the adaptability of the separable rules. The results show that
separable rules have their limitations; they cannot be rate adaptive,
which implies that information pooling is the key to achieve
adaptation. We then consider in Section \ref{oracle.sec} adaptive block
thresholding estimators through ideal adaptation with oracle.

\subsection{Adaptability of Separable Rules}
\label{separable.sec}

As discussed in Section \ref{minimax.sec}, Zhang (\citeyear{Zha05}) showed that
separable rules are asymptotically minimax over any given Besov ball
$\Bb$. Hence, from a minimax point of view there is little to gain by
looking beyond the separable rules when the parameters $(\al, p, q)$
are fully specified. A natural question is whether separable rules can
achieve the minimax rate of convergence simultaneously over a
collection of Besov balls. To answer this question, we begin with a
simple version of the adaptation problem by considering only two Besov
balls. Let $B^{\al_1}_{p_1, q_1}(M_1)$ and $B^{\al_2}_{p_2, q_2}(M_2)$
be two Besov balls with $\al_1 \neq\al_2$. We call an estimator $\de$
rate-adaptive over the two Besov balls if $\de$ attains the minimax
rate simultaneously over both of them, that is,
%
%
\begin{eqnarray}\label{rate.adaptive}
&&\max_{i =1, 2} \mathop{\overline{\lim}}_{n\goto\infty}
n^{2\al_i/(1+2\al_i)}\nonumber\\ [-8pt]\\ [-8pt]
&&\hphantom{\max_{i =1, 2}}{}\cdot \sup_{\theta\in B^{\al_i}_{p_i, q_i} (M_i)}
E\|\de- \theta\|_2^2 < \infty.\nonumber
\end{eqnarray}
The question is: can
(\ref{rate.adaptive}) be achieved by a separable rule? To answer the
question, Cai (\citeyear{Cai08}) showed that separable rules are ``inflexible'':
any rate-opti\-mal separable rule over a Besov ball $\Bb$ must have a
``flat'' rate of convergence everywhere in\break $\Bb$. If a separable rule
$\de$ satisfies
\[
\sup_{\theta\in\Bb} E\|\de-\theta\|_2^2 \le C n^{-2\al/(1+2\al)}
\]
for some constant $C > 0$, then for any given $\theta\in
B^\al_{p,q}(M)$,
%
%
\begin{eqnarray}\label{flat.rate}
0&<& \mathop{\underline{\lim}}_{n \goto
\infty}
n^{2\al/(1+2\al)} E\|\de-\theta\|_2^2 \nonumber\\ [-8pt]\\ [-8pt]
&\leq&\mathop{\overline{\lim
}}_{n\goto\infty}
n^{2\al/(1+2\al)} E\|\de-\theta\|_2^2 < \infty.\nonumber
\end{eqnarray}
That is, $\de$
must attain the exact same rate at every point $\theta\in\Bb$. This is
not the case for nonseparable rules. Indeed, there exist estimators
that converge faster than the minimax rate at every point in $\Bb$.\vspace*{2pt} See
Brown, Low and Zhao (\citeyear{BroLowZha97}), Zhang (\citeyear{Zha05}) and Cai (\citeyear{Cai08}). As a direct
consequence of the inflexibility of the separable rules, they are
necessarily not rate-adaptive. That is, if $\al_1 \neq\al_2$, then
%
%
\begin{eqnarray} \label{sep.cor.eq}
&&\max_{i =1, 2} \mathop{\overline{\lim}}_{n\goto\infty}
n^{2\al_i/(1+2\al_i)}\nonumber\\ [-8pt]\\ [-8pt]
&&\hphantom{\max_{i =1, 2}}{}\cdot\inf_{\de\in\mathcal{S}} \sup_{\theta\in B^{\al_i}_{p_i, q_i}
(M_i)} E\|
\de-
\theta\|_2^2 = \infty. \nonumber
\end{eqnarray}

The lack of adaptability of separable rules is close\-ly connected
to
superefficiency in the classical univariate normal\vadjust{\goodbreak} mean problem. It is
well known that if an estimator of a univariate normal mean is
superefficient at a point it must pay for the superefficiency by being
subefficient in a neighborhood of that point. The Hodges estimator is
an example of such estimators. See Le Cam (\citeyear{LeC53}) and Brown and Low
(\citeyear{BroLow96N2}).

Under the sequence model (\ref{seq.model}), the minimax rate of
convergence over the Besov ball $\Bb$ is\break $n^{-2\al/(1+2\al)}$. We call
an estimator $\de$ \textit{superefficient} at a fixed point $\theta\in
\Bb$ if
\[
n^{2\al/(1+2\al)}E_\theta\|\de-\theta\|^2_2\goto0.
\]
A heuristic proof of (\ref{flat.rate}) sheds light on the cause of the
lack of adaptability for separable rules. Let $\de= (\de_{j,k})$ be a
minimax rate-optimal separable rule over $B^\al_{p,q}(M)$. Then
individually each $\de_{j,k}$ can be regarded as an estimator in a
univariate normal mean problem. If $\de$ is superefficient at some
$\theta^*\in B^\al_{p,q}(M)$, then, as a univariate normal mean
problem, ma\-ny~$\de_{j,k}$ are superefficient at~$\theta^*_{j,k}$ and,
thus, each of these~$\de_{j,k}$ must be penalized in a subefficient
neighborhood of~$\theta^*_{j,k}$. There exists some $\theta' \in\Bb$
with coordinates~$\theta'_{j,k}$ in those subefficient neighborhoods
of $\theta^*_{j,k}$. As a\vspace*{2pt} consequence of~$\de$ being superefficient at
$\theta^*$, $\de$ is subefficient at~$\theta'$ relative to the minimax
risk over $\Bb$. This contradicts the assumption that~$\de$ is
rate-optimal uniformly over $B^\al_{p,q}(M)$. A~rigorous argument can
be found in Cai (\citeyear{Cai08}). The main reason this phenomenon occurs is that
separable rules estimate each coordinate~$\theta_{j,k}$ based solely on
an individual observation~$y_{j,k}$. Estimation accuracy can be
improved by pooling information on different coordinates to make more
informative and accurate decisions.

Equation (\ref{sep.cor.eq}) shows that separable rules need to pay a
price for adaptation. The minimum cost of~ada\-ptation for the separable
rules is at least a logarith\-mic factor. Suppose $\al_1 > \al_2$. If a
separable rule $\de$ attains the minimax rate $n^{2\al_1/(1+2\al_1)}$
over $B^{\al_1}_{p_1, q_1}(M_1)$, then
%
%
\begin{eqnarray}\label{sep.l.b}
&&\mathop{\underline{\lim}}_{n\goto\infty}
\biggl(\frac{n}{\log n}\biggr)^{2\al_2/(1+2\al_2)}\nonumber\\ [-8pt]\\ [-8pt]
&&\quad{}\cdot \sup_{\theta
\in
B^{\al_2}_{p_2, q_2}(M_2)} E\|\de- \theta\|_2^2 > 0.\nonumber
\end{eqnarray}
This lower bound bears a strong similarity to the problem of
adaptive estimation of a function at\break a~point. See Section
\ref{pointwise.sec}.

The lower bound (\ref{sep.l.b}) can indeed be attained by a~separable
rule. The well-known VisuShrink estimator of Donoho and
Johnstone\vadjust{\goodbreak}
(\citeyear{DonJoh94}) adaptively achieves within a logarithmic factor of the minimax
risk. It is thus optimal among separable rules in the sense that it
attains the lower bound on the adaptive convergence rate within this
class of estimators.

To motivate the VisuShrink estimator, we begin with the classical
multivariate normal mean model~(\ref{normal.model0}) and outline an
oracle approach developed in Donoho and Johnstone (\citeyear{DonJoh94}). Suppose we
wish to estimate $\theta= (\theta_1, \ldots, \theta_m)$ based on the
observations $x = (x_1,\break \ldots, x_m)$ in (\ref{normal.model0}) under
mean squared error~(\ref{risk}).

In the discussion that follows, we focus on the separable rules. An
ideal separable ``estimator'' $\hat\theta^{\mathrm{ideal}}$ would estimate
$\theta_i$ by $x_i$ when $\theta_i^2 > \sigma^2$ and by $0$~other\-wise,
that is, $\hat\theta_i^{\mathrm{ideal}} = x_i I(\theta_i^2 > \sigma^2)$.
This \mbox{``estimator''} achieves ideal trade-off between variance and
squa\-red bias for each coordinate and attains the ideal~risk
%
%
\begin{equation}\label{DP.oracle.risk}
 R_{\mathrm{DP.oracle}}(\theta) = \frac{1}{m} \sum_{i =
1}^m (\theta_i^2\wedge\sigma^2).
\end{equation}
Since the ``estimator'' $\hat
\theta^{\mathrm{ideal}}$ requires the knowledge of the unknown $\theta$, it is
not a true statistical estimator. The ideal risk (\ref{DP.oracle.risk})
is unattainable in practice, but it does provide a useful benchmark. To
mimic the performance of the ideal ``estimator'' $\hat\theta^{\mathrm{ideal}}$,
Do\-noho and Johnstone (\citeyear{DonJoh94}) proposed the soft threshold estimator
%
%
\begin{equation}
\hat\theta_i^* = \sgn(x_i) (|x_i| - \tau)_+,\label{soft}
\end{equation}
with
$\tau= \sigma\sqrt{2 \log m}$, and showed the following Oracle
Inequality:
%
%
\begin{eqnarray}\label{DP}
&&R(\hat\theta^*, \theta)\nonumber\\
&&\quad \leq(2 \log m + 1)
[R_{\mathrm{DP.oracle}}(\theta) + \sigma^2/m],\\
\eqntext{\mbox{ for all $\theta
\in\RR^m$.}}
\end{eqnarray}
Hence, the soft threshold estimator
$\hat
\theta^*$ comes within a logarithmic factor of the ideal risk for all
$\theta\in\RR^m$. Moreover, the factor $2 \log m$ in the Oracle
Inequali\-ty~(\ref{DP}) is asymptotically sharp in the following sense:
%
%
\begin{eqnarray} \label{minimax0}
&&\inf_{\hat\theta} \sup_{\theta\in\RR^m}
\frac{E\|\hat\theta- \theta\|_2^2}
{\sigma^2 + \sum_{i = 1}^m \min(\theta_i^2, \sigma^2)}\nonumber\\ [-8pt]\\ [-8pt]
&&\quad = 2 \log m
\bigl(1+o(1)\bigr),\quad
 m \goto\infty.\nonumber
\end{eqnarray}
A similar result
to (\ref{minimax0}) is given in Foster and George (\citeyear{FosGeo94}) in the linear
regression setting.

In the setting of the Gaussian sequence mo\-del~(\ref{seq.model}),
VisuShrink is defined as
%
%
\begin{equation}\label{visu}
\hspace*{20pt}\hat\theta_{j,k} = \cases{
\sgn(y_{j,k}) \bigl(|y_{j,k}| -\sqrt{2 n^{-1}\log n}\bigr)_+,\vspace*{1pt}\cr
\hspace*{22pt} \mbox{if } j < J,\cr
0,\quad\mbox{if  $j \geq J$},}
\end{equation}
where $J=\lfloor\log_2 n\rfloor$. The
VisuShrink estimator adaptively achieves the rate of convergence $(\log
n/\break n)^{2\al/(1+2\al)}$ over the Besov balls $\Bb$ (Donoho et al., \citeyear{Donetal95}).
That is,
%
%
\begin{equation}
\hspace*{15pt}\sup_{\theta\in\Bb} E\|\hat\theta-\theta\|_2^2
\leq C
\biggl(\frac{\log n}{n}\biggr)^{2\alpha/(1 + 2\alpha)},
\end{equation}
where
$C>0$ is a constant not depending on $n$. In light of the lower bound
(\ref{sep.l.b}), VisuShrink is thus optimal within the class of
separable rules.

\subsection{Block Thresholding via Ideal Adaptation~with~Oracle}
\label{oracle.sec}

The results in Section \ref{separable.sec} show that information
pooling is a necessity for achieving full adaptation. Block
thresholding, which estimates the coordinates in groups rather than
individually, provides a conve\-nient and effective tool for information
pooling. Block thresholding increases estimation precision and\break achie\-ves
adaptivity by utilizing information about neighboring coordinates. The
degree of adaptivity, however, depends on the choice of block size and
threshold level.

We study block thresholding rules via the approach of ideal adaptation
with an oracle. The main ideas of the oracle approach have been
outlined at the end of Section \ref{separable.sec} in developing the
VisuShrink estimator. An oracle does not reveal the true estimand, but
provides the ideal choice within a given class of estimators. The
oracle ``estimator'' is typically not a true statistical estimator, as
it may depend on the unknown parameter. It represents an ideal for a
particular estimation method. The goal of ideal adaptation is to derive
true statistical estimators which can essentially mimic the performance
of an oracle.

The soft threshold estimator (\ref{soft}) estimates coordinates
individually without using information about other coordinates. As we
have shown in Section \ref{separable.sec}, such a separable rule is not
optimal for adaptive estimation. We thus consider a more general class
of estimators, the block projection (BP) estimators, which use
information about neighboring coordina\-tes by thresholding observations
in groups. Simultaneous decisions are made to retain or discard all the
coordinates within the same group.

We again begin with the finite-dimensional multivariate normal mean
model (\ref{normal.model0}). We wish to estimate the mean $\theta=
(\theta_1, \ldots, \theta_m)$ based on the observations $x = (x_1,
\ldots, x_m)$ in (\ref{normal.model0}) under the mean squared error
(\ref{risk}). Let $B_1, B_2, \ldots, B_N$ be a partition of the
index
set $\{1, \ldots, m\}$ with each $B_i$ of size $L$ (for convenience,\vadjust{\goodbreak}
we assume that the sample size $m$ is divisible by the block size $L$).
Let $\mathcal{H}$ be a subset of the block indices $\{1, \ldots, N\}
$. A
block projection estimator $\hat\theta(\mathcal{H}) $ is defined as
%
%
\begin{eqnarray}\label{BP.est}
\hat\theta_{B_j}(\mathcal{H}) &=& x_{B_j}\quad \mbox{if $j \in
\mathcal{H}$}
\quad\mbox{and}\nonumber\\ [-8pt]\\ [-8pt]
\hat\theta_{B_j}(\mathcal{H}) &=& 0\quad \mbox{if
$j \notin\mathcal{H}$},\nonumber
\end{eqnarray}
where
$x_{B_j}=(x_i)_{i\in
B_j}$. The risk of $\hat\theta(\mathcal{H})$ is
%
%
\begin{eqnarray}\label{risk11}
&&\hspace*{20pt}R(\hat\theta
(\mathcal{H}), \theta)\nonumber\\ [-5pt]\\ [-8pt]
&&\hspace*{20pt}\quad = \frac{1}{m} \sum_{j=1}^N \{L \sigma^2 I(j
\in\mathcal{H}) +
\|\theta_{B_j}\|^2_2 I(j \notin\mathcal{H})\}.\nonumber
\end{eqnarray}
Ideally, one would like to choose $\mathcal{H}$ to consist of blocks $j$
where $\|\theta_{B_j}\|^2_2 > L \sigma^2$. A BP oracle provides
exactly this side information $\mathcal{H}_* = \mathcal{H}_*(\theta)
= \{j\dvtx\break
\|\theta_{B_j}\|^2_2 > L \sigma^2\}$, which yields the ideal block
projection ``estimator'' $\hat\theta(\mathcal{H}_*)$ with $\hat
\theta_{B_j}(\mathcal{H}_*) = x_{B_j} I(j \in\mathcal{H}_*)$ with
the ideal
risk
%
%
\begin{eqnarray}\label{ideal.risk}
R_{\mathrm{BP.oracle}}(\theta, L) &=& \inf_\mathcal{H} \frac
{1}{m}E\|
\hat
\theta(\mathcal{H}) - \theta\|_2^2 \nonumber\\ [-8pt]\\ [-8pt]
&=&\frac{1}{m} \sum^N_{j=1}
(\|\theta_{B_j}\|^2_2\wedge L \sigma^2).\nonumber
\end{eqnarray}

The ideal ``estimator'' $\hat\theta(\mathcal{H}_*)$ is not a true
statistical estimator. A natural goal is to construct an estimator
which can mimic the performance of the BP oracle.

Since Stein's 1956 seminar paper, many shrinkage estimators have been
developed in the multivariate normal decision theory. Among them, the
(positive part) James--Stein estimator is perhaps the best-known. Efron
and Morris (\citeyear{EfrMor73}) showed that the (positive part) James--Stein estimator
does more than just demonstrate the inadequacy of the maximum
likelihood estimator; it is a member of a class of good shrinkage
rules, all of which may be useful in different estimation problems.
Indeed, as we shall see below, blockwise James--Stein rules can
essentially mimic the performance of the BP oracle when the threshold
is properly chosen. For each block $B_j$ let $S_j^2 = \sum_{i\in B_j}
x_i^2$ and set
%
%
\begin{equation} \label{JS}
\hat\theta_{B_j}(L, \lam) = \biggl(1 - \frac{\lam L
\sigma^2}{S^2_j}\biggr)_+ x_{B_j}.
\end{equation}
Then the blockwise
James--Stein estimator satisfies the following BP Oracle Inequality:
%
%
\begin{eqnarray}\label{BP.oracle3}
&&\hspace*{15pt}R(\hat\theta(L, \lam), \theta)\nonumber\\ [-8pt]\\ [-8pt]
&&\hspace*{15pt}\quad \leq\lam
R_{\mathrm{BP.oracle}}(\theta,
L) + 4 \sigma^2 \cdot P(\chi^2_L > \lam L),\nonumber
\end{eqnarray}
where $\chi^2_L$ denotes a central chi-squared random variable with $L$
degrees of freedom.
\begin{remark}
When the block size $L=1$, the estimator
(\ref{JS}) becomes a
coordinatewise thresholding estimator. It is easy to show that with
the choice of $\lam= 2\log m$ the BP Oracle Inequality
(\ref{BP.oracle3}) is equivalent to the Oracle Inequality
(\ref{DP}) of Donoho and Johnstone (\citeyear{DonJoh94}). The resulting estimator
shares similar properties with the VisuShrink estimator. See Gao (\citeyear{Gao98}).
\end{remark}

\begin{remark}
Another special choice of block size is $L=L_* = \log m$. The
corresponding threshold is $\lam=\lam_*\equiv4.50524$ (the solution of
$\lam- \log\lam- 3 =0$). The pair $(L_*, \lam_*)$ is chosen so that
the corresponding estimator in the Gaussian sequence model is (near)
optimal. See the discussion below. In this case the BP Oracle
Inequality becomes
%
%
\begin{equation} \label{oracle0}
\hspace*{20pt}R(\hat\theta(L_*, \lam_*), \theta) \leq
\lam_* R_{\mathrm{BP.oracle}}(\theta, L_*) +\frac{2\sigma^2}{m}.
\end{equation}

Therefore, with block size $L_* = \log m$ and thresholding constant
$\lam_*= 4.50524$, the estimator comes essentially within a constant
factor of 4.50524 of the ideal risk. Note that this blockwise
James--Stein estimator is not minimax for a given block (since \mbox{$\lam_*
>2$}), but it is close to being minimax and $\lam_*= 4.50524$ is needed
for the optimal performance in the infinite-dimensional Gaussian
sequence model.
\end{remark}

\begin{remark}
Instead of the block projection estimators given in
(\ref{BP.est}), one can also consider the more general block linear
shrinkers: $\hat\theta_{B_j} = \gamma_j x_{B_j}, \gamma_j\in[0,
1].$ In the case of block projection, $\gamma_j\in\{0, 1\}.$ An oracle
would provide the ideal shrinkage factors $\gamma_j =
\|\theta_{B_j}\|_2^2/(\|\theta_{B_j}\|_2^2 + L \sigma^2),$ and the
ideal ``estimator'' has the risk
\[
R_{\mathrm{BLS.oracle}}(\theta, L) = \frac{1}{m} \sum_{j=1}^N
\frac{\|\theta_{B_j}\|_2^2 L \sigma^2}{\|\theta_{B_j}\|_2^2 + L
\sigma^2}.
\]

The blockwise James--Stein estimator (\ref{JS}) also mimics the
performance of the block linear shrinker oracle,
%
%
\begin{eqnarray}
&&\hspace*{15pt}R(\hat\theta(L,
\lam), \theta)\nonumber\\ [-8pt]\\ [-8pt]
&&\hspace*{15pt}\quad \leq2 \lam R_{\mathrm{BLS.oracle}}(\theta, L) + 4 \sigma^2
\cdot P(\chi^2_L > \lam L).\nonumber
\end{eqnarray}
\end{remark}

We now return to the Gaussian sequence mo\-del~(\ref{seq.model}) and
consider the BlockJS procedure introduced in Cai (\citeyear{Cai99}).
Let $J =[\log_2 n]$. Divide each resolution level $1\leq j < J$ into
nonoverlapping blocks of length $L =L_* = [\log n]$. (The coordinates
in the first few resolution levels are grouped into a single
block.)\vadjust{\goodbreak}
Let $b^j_i$ denote the set of indices of the coordinates in the $i$th
block at level $j$, that is,
\[
b^j_i = \{(j, k)\dvtx (i-1) L + 1 \leq k \leq i L\}.
\]
Set $S^2_{j,i} \equiv\sum_{k\in b^j_i} y_{j,k}^2$. We\vspace*{-2pt} then apply the
James--Stein shrinkage rule to each block $b^j_i$. For
$(j,k) \in b^j_i$,
%
%
\begin{equation} \label{seq.est}
\hat\theta_{j,k}^* =\cases{
\biggl(1 - \dfrac{\lam_* L n^{-1}}{S_{j,i}^2}\biggr)_+ y_{j,k}, \cr
\hspace*{22pt}\mbox{for $(j,k) \in b^j_i, j < J$},\cr
0,\quad \mbox{for $j \geq J$},
}
\end{equation}
where $\lam_*\equiv4.50524$ is the
solution of $\lam- \log\lam- 3 =0$. This threshold is derived based
on the tail probability of a chi-squared distribution. See Cai (\citeyear{Cai99}).

The BlockJS estimator (\ref{seq.est}) is adaptively within a~constant
factor of the minimax risk over all Besov balls $\Bb$ for $p \ge2$ and
is within a logarithmic factor of the minimax risk over Besov balls
$\Bb$ for $p < 2$,
%
%
\begin{eqnarray}
&&\hspace*{20pt}\sup_{\theta\in\Bb} E\|\hat\theta^*-\theta\|_2^2 \nonumber\\ [-8pt]\\ [-8pt]
&&\hspace*{20pt}\quad \leq\cases{
C n^{-2\alpha/(1 + 2\alpha)} \cr
\quad \mbox{for $p \geq2$}\vspace*{2pt}\cr
C n^{-2 \alpha/(1 + 2 \alpha)} (\log n)^{(2/p -1)/(1 + 2
\alpha)} \cr
\quad \mbox{for $p < 2$ and $\al p \geq1$.}\nonumber
}
\end{eqnarray}
The block size and threshold level play important roles in the
performance of a block thresholding estimator. The block size $L_* =
\log n$ and threshold $\lam_* = 4.50524$ are shown in Cai (\citeyear{Cai99}) to be
optimal in the sense that the resulting BlockJS estimator is both
globally and locally adaptive. The extra logarithmic factor in the case
of $p < 2$ is unavoidable for any block thresholding estimators with
fixed block size and threshold.

Adaptation can be achieved through empirically selecting the block size
and threshold at each resolution level by minimizing Stein's Unbiased
Risk Estimate (Cai and Zhou, \citeyear{CaiZho09}). Let $y_{j.} = (y_{j,1}, \ldots, y_{j,
2^j})$. Since the positive part James--Stein estimator (\ref{JS}) is
weakly differentiable, Stein's formula (Stein, \citeyear{Ste81}) for unbiased
estimate of risk shows that
\begin{eqnarray*}
&&\operatorname{SURE}(y_{j.}, L, \lam)\\
&&\quad \equiv2^j + \sum_{i}\frac{\lam^2L^2 - 2
\lam L(L-2)}{S_{(jb)}^2}\cdot I(S_{j,i}^2> \lam L)\\
&&\qquad{} + (S^2_{j,i} - 2
L)\cdot I(S^2_{j,i}\leq\lam L)
\end{eqnarray*}
is an unbiased estimate of the risk at level $j$. Choose the
level-dependent block size $L_j$ and threshold\vadjust{\goodbreak} $\lam_j$ to be the
minimizer of $\operatorname{SURE}$:
\[
(L_j, \lam_j) = \operatorname{arg\,min}\limits_{L, \lam} \operatorname{SURE}(y_{j.}, L,
\lam).
\]

The resulting estimator, called SureBlock, auto\-matically adapts to the
sparsity of the underlying~se\-quence~$\theta$. In particular, the
estimator is sharp adapti\-ve over all Besov balls $B_{2, 2}^\alpha(M)$
and simultaneously achieves within a factor of 1.25 of the minimax
risk over Besov\vspace*{1pt} balls $\Bb$ for all $p\ge2$, $q\ge2$. At the\vspace*{1pt} same
time the SureBlock estimator achieves adapti\-vely within a~constant
factor of the minimax risk over a~wide collection of Besov balls $\Bb$
in the ``sparse case'' $p < 2$. These properties are not shared
simultaneous\-ly by other commonly used thresholding procedures such as
VisuShrink (Donoho and\break Johnstone, \citeyear{DonJoh94}), SureShrink (Donoho and
Johnstone, \citeyear{DonJoh95}) or BlockJS (Cai, \citeyear{Cai99}).

\subsection{Discussion}


The idea of block thresholding can be traced back to Efromovich (\citeyear{Efr85})
in estimation using the trigonometric basis. A similar construction was
used in\break Brown, Low and Zhao (\citeyear{BroLowZha97}) to produce superefficient
estimators. In the context of wavelet estimation, global level-by-level
thresholding was discussed in Donoho and Johnstone (\citeyear{DonJoh95}) for
regression and in Kerkyacharian, Picard and Tribouley (\citeyear{KerPicTri96}) for
density estimation. Cavalier and Tsybakov (\citeyear{CavTsy02}) and Cavalier et al. (\citeyear{Cavetal03}) and Cai, Low and Zhao (\citeyear{CaiLowZha09}) used weakly
geometrically growing block size for sharp adaptation over ellipsoids.
But these block thresholding methods are not local, they essentially
adaptively mimic the performance of the ideal linear estimator. Because
of the serious limitations of the linear procedures for estimating
spatially inhomogeneous functions discussed at the end of Section
\ref{linear.sec}, these estimators do not enjoy a~high degree of
spatial adaptivity. In particular, these estimators do not perform well
over parameter spaces which are not quadratically convex such as Besov
balls $\Bb$ with $p < 2$.\vspace*{1pt}

Hall, Kerkyacharian and Picard (\citeyear{HalKerPic98}, \citeyear{HalKerPic99}) introduced a local
blockwise hard thresholding procedure for density estimation and
nonparametric regression with a block size of the order $(\log n)^{2}$
whe\-re~$n$ is the sample size. Cai and Silverman (\citeyear{CaiSil01}) considered
overlapping block thresholding estimators. Block thresholding is a
widely applicable technique.
Cai and Low (\citeyear{CaiLow05N2}, \citeyear{CaiLow06N2}) use block thresholding procedures for
minimax as well as optimal adaptive estimation of a quadratic
functional and Cai and Low (\citeyear{CaiLow06N1}) used a block thresholding me\-thod for
the construction of adaptive confidence balls.\

We have focused the discussion on\vspace*{-2pt} blockwise Ja\-mes--Stein procedures
because of their simplicity. In addition to the James--Stein rule,
through block thresholding, many other shrinkage rules developed in the
classical normal decision theory can be applied as well. For example,
estimators of the forms
\[
\hat\theta= [1 - \lam_1 \sigma^2/(\lam_2 + S^2)]_+ y
\quad\mbox{or}\quad\hat\theta= [1 - c(S^2)]_+ y,
\]
where $S^2 =\|y\|_2^2$ and $c(\cdot)$ is a suitably chosen function,
can also be used. Besides block thresholding, the empirical Bayes
method is another natural choice for information pooling and for
constructing adaptive procedures. See Johnstone and Silverman (\citeyear{JohSil05})
and Zhang (\citeyear{Zha05}). In particular, Zhang (\citeyear{Zha05}) presented a class of
general empirical Bayes estimators that are adaptively sharp minimax
over a large collection of Besov balls. Other methods such as choosing
a threshold by controlling the false discovery rate can also be used.
See Abramovich et al. (\citeyear{Abretal06}).

\section{Minimax and Adaptive Estimation under~Pointwise Loss}
\label{pointwise.sec}

So far the focus has been on the minimax and adaptive estimation under
the global MISE risk (\ref{g.risk}). For functions of spatial
inhomogeneity, the local\break smoothness of the functions varies
significantly from point to point and global risk measures such as
(\ref{g.risk}) cannot wholly reflect the local performance of an
estimator. The most commonly used measure of local accuracy is
pointwise squared error loss. While the minimax theory under the
pointwise loss is similar to that for the global loss, the adaptation
theories for the two losses are significantly different. Under the
local loss it is often the case that sharp adaptation is not possible
and a penalty, usually a logarithmic factor, must be paid for not
knowing the smoothness. Estimation under the pointwise risk
(\ref{l.risk}) is a~special case of estimating a linear functional
$T(f)$. A general theory for estimating linear functionals has been
developed in the literature. In this section we shall first focus on
estimating a function under the pointwise risk and present a concise
account of both the minimax and adaptation results. The related minimax
and adaptation theory for estimating a~general linear functional is
discussed in Section \ref{l.functional.sec}.

We shall return to the white noise model (\ref{w.model}) and consider
estimation under pointwise squared error risk
%
%
\begin{equation}\label{l.risk}
R(\hat f, f; t_0) =
E_f \bigl(\hat f(t_0) - f(t_0)\bigr)^2,
\end{equation}
where $t_0\in(0,
1)$ is any fixed point. For a given parameter space $\mathcal{F}$, the
difficulty of the estimation problem is measured by the minimax risk
%
%
\begin{equation} \label{minimax.risk}
R_n^*(\mathcal{F}; t_0) = \inf_{\hat f} \sup
_{f\in\mathcal
{F}} E_f \bigl(\hat
f(t_0) - f(t_0)\bigr)^2.
\end{equation}
Several methods have been
developed to study the minimax estimation problem. These include
modu\-lus of continuity, metric entropy, information inequa\-lity,
renormalization and constrained risk inequality. See, for example,
Farrell (\citeyear{Far72}), Hasminskii (\citeyear{Has79}), Stone (\citeyear{Sto80}), Ibragimov and
Hasminskii (\citeyear{IbrHas84}), Do\-noho and Liu (\citeyear{DonLiu91}), Brown and Low (\citeyear{BroLow91}), Low
(\citeyear{Low92}), Donoho and Low (\citeyear{DonLow92}) and Birg\'{e} and Massart (\citeyear{BirMas95}). For
example, the minimax risk over any convex parameter space can be
characterized, up to a small constant factor, in terms of the modulus
of continuity. For estimation over the Besov balls, the minimax rate
of convergence of the pointwise risk is derived in Cai (\citeyear{Cai03}) using a
constrained risk inequality. It is shown that the minimax risk
satisfies
%
%
\begin{equation}
R_n^*(\Bb; t_0) \asymp n^{-2\nu/(1 + 2\nu)},
\end{equation}
where $\nu= \al- \frac{1}{p}$. Unlike the minimax rate of convergence
under the global risk, the local minimax rate of convergence depends on
the parameter $p$ as well. Minimax rate optimal estimators can be
constructed using wavelet thresholding.

The behavior of the estimators which are minimax rate optimal under the
pointwise risk is quite different from that of rate optimal estimators
under the global MISE risk. It is shown in Cai (\citeyear{Cai03}) that if an
estimator $\hat f$ attains the minimax rate of convergence over a Besov
ball $\Bb$, then it must attain the same ``flat'' rate at every $f$ in
the parameter space; superefficiency is not possible for rate optimal
estimators. That is, if
%
%
\begin{eqnarray}\label{rate-optimal}
&&\mathop{\overline{\lim}}_{n\goto\infty} n^{2 \nu/(1+
2\nu)}\nonumber\\ [-8pt]\\ [-8pt]
&&\quad{}\cdot \sup_{f\in\Bb} E_f \bigl(\hat f(t_0) - f(t_0)\bigr)^2 < \infty,\nonumber
\end{eqnarray}
then the estimator $\hat f$ must also satisfy
%
%
\begin{equation}\label{flat-rate}
\mathop{\underline{\lim}}_{n\goto\infty} n^{2 \nu/(1+
2\nu)} E_f\bigl(\hat
f(t_0) -
f(t_0)\bigr)^2 > 0
\end{equation}
for any fixed $f\in\Bb$. In
contrast, under the~glo\-bal MISE risk, rate-optimal estimators
over
$\Bb$ can achieve a much faster rate at some parameter points. Indeed,
it is possible to have estimators which converge at a rate faster than
the minimax rate at every fixed function in $\Bb$; see\vspace*{2pt} Brown, Low and
Zhao (\citeyear{BroLowZha97}), Zhang (\citeyear{Zha05}) and Cai
(\citeyear{Cai08}).\vadjust{\goodbreak}

Pioneering work on adaptive estimation under the pointwise risk began
with Lepski (\citeyear{Lep90}). This work focused on Lipschitz balls and showed
that it is impossible to achieve complete adaptation for free when the
smoothness parameter is unknown. One must pay a price for adaptation.
Lepski (\citeyear{Lep90}) and Brown and Low (\citeyear{BroLow96N2}) showed that the cost of
adaptation is at least a logarithmic factor even when the smoothness
parameter is known to be one of two values. The case of the Sobolev
balls was investigated by Tsybakov (\citeyear{Tsy98}). Cai (\citeyear{Cai03}) considered
adaptation over Besov balls.

The inflexibility of the minimax rate optimal estimators has direct
consequence for adaptive estimation over Besov balls under the
pointwise loss. Adaptation for free is only possible if the rates of
convergence over the collection of the Besov balls are the same, that
is, $\nu= \al- {1\over p}$ is a fixed constant for all Besov balls in
the collection. Otherwise, a~penalty must be paid for adaptation, even
over two Besov balls $B^{\al_i}_{p_i,q_i}(M_i)$, $i=1, 2$. Let $\nu_i
\equiv\al_i - 1/p_i$ for $i = 1, 2$ and suppose $\nu_1 > \nu_2> 0$. If
an estimator $\hat f$ attains a rate of $n^\rho$ over $\Bbi$ with
$\rho
> 2\nu_2/(1+2\nu_2)$, in particular, if $\hat f$ is rate-optimal over
$\Bbi$, then
\begin{eqnarray*}
&&\mathop{\underline{\lim}}_{n\goto\infty} \biggl(\frac{n}{\log n}\biggr)^{2 \nu_2/(1
+ 2\nu_2)}\\
&&\quad{}\cdot \sup_{f\in\Bbii} E_f \bigl(\hat f(t_0) - f(t_0)\bigr)^2 > 0.
\end{eqnarray*}
Therefore, the minimum cost for adaptation is at least a logarithmic
factor. Furthermore, the rate $(n/\break\log n)^{2 \nu/(1 + 2\nu)}$ can be
adaptively attained, for example, by the VisuShrink estimator of Donoho
and Johnstone (\citeyear{DonJoh94}) and the BlockJS estimator discussed in Section~\ref{oracle.sec}.
See Cai (\citeyear{Cai03}).

\begin{remark*}
We have focused on adaptation over different parameter
spaces under a given loss. There is another type of adaptation problem
which can be termed as loss adaptation: given a fixed parameter space,
is it possible to construct an estimator that adapts to the loss
function in the sense that the estimator is optimal both locally and
globally? This problem was considered in Cai, Low and Zhao (\citeyear{CaiLowZha07}). It
was shown that it is impossible for any estimator to simultaneously
attain the global minimax rate of convergence and the local minimax
rate at every point when the global and local minimax rates are
different. The minimum penalty for a global rate-optimal estimator is a
logarithmic factor in terms of the maximum pointwise risk over $\Bb$.
The wavelet\vspace*{1.5pt} thresholding estimator with coefficients estimated by
(\ref{term.est}) is optimally loss adaptive in this sense.
\end{remark*}

\subsection{Discussion on Estimation of~Linear~Functionals}
\label{l.functional.sec}

The problem of estimating a function under the pointwise risk
(\ref{l.risk}) is a special case of estimating a linear functional
$T(f)$. For a given linear functional $T$ and a parameter space
$\mathcal{F}$
define the linear minimax risk $R_n^L(\mathcal{F}, T)$ and minimax risk
$R_n^*(\mathcal{F}, T)$, respectively, by
\begin{eqnarray*}
R_n^L(\mathcal{F}, T) &=& \inf_{\hat T\ \mathrm{linear}} \sup_{f\in
\mathcal{F}}
E_f\bigl(\hat
T - T(f)\bigr)^2\quad \mbox{and}\\
 R_n^*(\mathcal{F}, T) &=& \inf_{\hat T}
\sup_{f\in\mathcal{F}} E_f\bigl(\hat T - T(f)\bigr)^2.
\end{eqnarray*}
The minimax theory for estimating a linear functional~$T$ over a convex
parameter space has been well developed. See, for example, Ibragimov
and Hasminskii (\citeyear{IbrHas84}), Donoho and Liu (\citeyear{DonLiu91}) and Donoho (\citeyear{Don94}). In
particular, the properties of the minimax linear estimators can be
described precisely and the linear minimax risk $R_n^L(\mathcal{F},
T)$ is
within a small constant factor (${\le}1.25$) of the minimax risk
$R_n^*(\mathcal{F}, T)$, that is,
\[
R^L_n(\mathcal{F}, T) \le\mu^* R^*_n(\mathcal{F}, T) \le1.25
R^*_n(\mathcal{F}, T),
\]
where $\mu^*$ is the Ibragimov--Hasminskii constant given in
(\ref{IH.constant}). A fundamental quantity which captures the
difficulty of the estimation problem in this setting is the modulus of
continuity
%
%
\begin{eqnarray}\label{modulus}
&&\om(\ep, \mathcal{F})\nonumber\\
&&\quad = \sup\{
|T(g) - T(f)|\dvtx
\|g - f\|_2\leq\ep,\\
&&\hspace*{140pt}f, g \in\mathcal{F}\}.\nonumber
\end{eqnarray}
For example,
the linear
minimax risk is given by
%
%
\begin{equation}\label{BB}
R_n^L(\mathcal{F}, T) = \sup_{\epsilon>
0}\frac{\omega^2(\epsilon,\mathcal{F})}{4 + n \epsilon^2}
\end{equation}
and satisfies
\begin{eqnarray*}
\tfrac{1}{5}\omega^2(n^{-1/2}, \mathcal{F}) &\le& R_n^*(\mathcal
{F}, T) \le
R_n^L(\mathcal{F}, T)\\
 &\le&\omega^2(n^{-1/2}, \mathcal{F}).
\end{eqnarray*}
See Ibragimov and Hasminskii (\citeyear{IbrHas84}) and Donoho and Liu (\citeyear{DonLiu91}).

In most common cases when estimating a linear functional over convex
parameter spaces the modulus is H\"{o}lderian,
%
%
\begin{equation}\om(\ep,
\mathcal{F}) =
C\epsilon^{q(\mathcal{F})}\bigl(1 + o(1)\bigr).
\end{equation}
In this case the exponent
$q(\mathcal{F})$
determines the minimax rate of convergence. Hence, the rate of
convergence is captured by the geometric quantity $\om$. Furthermore,
Donoho and Liu (\citeyear{DonLiu91}) showed that the modulus can be used to give a
recipe for constructing the minimax linear estimator. A key step in
this analysis is to show that the difficulty for linear estimators over
a convex parameter space is in fact equal to the difficulty for linear
estimators of the hardest one-dimensional subproblem. This problem is
again closely connected to the problem of estimating a one-dimensional
bounded normal mean discussed in Section \ref{linear.sec}. Cai and Low
(\citeyear{CaiLow04N1}) extended the minimax theory for estimating linear functionals to
nonconvex parameter spaces. It is shown that in this setting while the
minimax rate of convergence is still determined by the modulus of
continuity, the linear minimax risk can be arbitrarily far from the
minimax risk. In fact, even if the parameter space is only a union of
two convex sets, it is possible that the maximum risk of the best
linear estimator does not even converge even though the minimax risk
converges quickly. This shows that linear estimators have serious
limitations when the parameter space is not convex.

The adaptation theory for estimating linear functionals is less well
developed. As mentioned earlier, Lepski (\citeyear{Lep90}) was the first to give
examples which demonstrated that rate optimal adaptation over a~collection of Lipschitz classes is not possible when estimating the
function at a point. Efromovich and Low (\citeyear{EfrLow94}) showed that this
phenomena is true in general over a~collection of nested symmetric
sets.
On the other hand, the goal of rate adaptive estimation of linear
functionals can sometimes be realized. When the minimax rates over each
parameter space are slower than any algebraic rate, Cai and Low (\citeyear{CaiLow03})
have given examples of nested symmetric sets where sharp adaptive
estimators can be constructed. In addition, when the parameter spaces
are not symmetric, there are also examples where rate adaptive
estimators can be constructed. See Efromovich (\citeyear{Efr97N1}, \citeyear{Efr97N2}, \citeyear{Efr00}), Lepski
and Levit (\citeyear{LepLev98}), Efromovich and Koltchinskii (\citeyear{EfrKol01}) and Kang and Low
(\citeyear{LowKan02}).

A general adaptation theory for estimating linear functionals is
given\vadjust{\goodbreak}
in Cai and Low (\citeyear{CaiLow05N1}). This theory gives a geometric characterization
of the adaptation problem analogous to that given by Donoho (\citeyear{Don94}) for
minimax theory. This theory describes exactly when rate adaptive
estimators exist, and when they do not exist the theory provides a
general construction of estimators with minimum adaptation cost.

It is shown that two geometric quantities, a between class modulus of
continuity and an ordered modulus of continuity, play a fundamental
role in the adaptation theory. The between class modulus of
continuity, defined by
%
%
\begin{eqnarray}\label{b.modulus}
&&\om_+(\ep, \mathcal{F}_1,
\mathcal{F}_2)\nonumber\\
&&\quad = \sup\{ |T(g) - T(f)| \dvtx \|g - f \|_2 \le\ep;\\
&&\hspace*{110pt}f
\in\mathcal{F}_1,
g \in\mathcal{F}_2\},\nonumber
\end{eqnarray}
captures the degree of adaptability over
two convex
parameter spaces in the same way that the usual modulus of continuity
used by Donoho and Liu (\citeyear{DonLiu91}) and Donoho (\citeyear{Don94}) captures the minimax
difficulty of estimation over a single convex parameter space. The
ordered modulus of continuity, given by
%
%
\begin{eqnarray}\label{o.modulus}
&&\om(\ep,
\mathcal{F}_1, \mathcal{F}_2)\nonumber\\
&&\quad = \sup\{ T(g) - T(f) \dvtx \|g - f \|
_2 \le\ep;\\
&&\hspace*{105pt}f
\in
\mathcal{F}_1, g \in\mathcal{F}_2\},\nonumber
\end{eqnarray}
is instrumental in the
construction of
adaptive estimators with minimum adaptation cost.

The theory shows that there are three main cases in terms of the cost
of adaptation. In the first case, the cost of adaptation is a
logarithmic factor of $n$. This is the case for estimating a function
at a point over Lipschitz balls. In the second case sharp adaptation is
possible as in the examples considered in Lepski and Levit (\citeyear{LepLev98}) and
Cai and Low (\citeyear{CaiLow03}). This is also the case when estimating a convex or
some other shape constrained function at a point. More dramatically, in
the third case the cost of adaptation is much greater than in the first
case. The cost of adaptation in this case is a power of~$n$.

\section{Minimax and Adaptive Confidence~Intervals}
\label{CI.sec}

The construction of confidence sets is an important part of statistical
inference. As mentioned in the introduction, there are several types of
nonparametric confidence sets including confidence intervals,
confidence bands and confidence balls. For example, Li (\citeyear{Li89}), Beran
and D\"{u}mbgen (\citeyear{BerDum98}), Genovese and Wasserman (\citeyear{GenWas05}), Cai and Low
(\citeyear{CaiLow06N1}) and Robins and van der Vaart (\citeyear{Robvan06}) have constructed confidence
balls with near optimal variable radius which also guarantee coverage
probability. Adaptive confidence bands have been constructed in the
special case of shape restricted functions. See Hengartner and Stark
(\citeyear{HenSta95}) and D\"{u}mbgen (\citeyear{Dum98}). See also Genovese and Wasserman (\citeyear{GenWas08}).

In this section we shall focus our discussion on pointwise confidence
intervals for a function. Similar to estimation under the pointwise
risk, this problem is a special case of confidence intervals for linear
functionals. Both minimax theory and adaptation theory for confidence
intervals of linear functionals have been developed. In this section we
shall first discuss the general theory and then use confidence
intervals for a function at a point as examples. Again, we will mainly
use the Besov balls $\Bb$ as the examples. The usual cases of
H\"{o}lder balls and Sobolev balls follow by taking $p=q=\infty$ and
$p=q=2$, respectively.

For any confidence interval there are two interrelated issues which
need to be considered together, coverage probability and the expected
length. A minimax theory for confidence intervals of linear functionals
was given in Donoho (\citeyear{Don94}) for convex parameter spaces. In this setting
the goal is to construct confidence intervals with a prespecified
coverage probability which minimizes the expected length of the
interval. Write $\si_{\gamma, \mathcal{F}}$ for the collection of all
confidence intervals which cover the linear functio\-nal~$T(f)$ with
minimum coverage probability of $1-\gamma$ over the parameter space
$\mathcal{F}$. Denote by
\[
L(\mathit{CI}, \mathcal{F}) = \sup_{f \in\mathcal{F}}E_f(L(\mathit{CI}))
\]
the maximum expected length of a confidence interval $\mathit{CI}$ over
$\mathcal{F}$
where $L(\mathit{CI})$ is the length of $\mathit{CI}$. The benchmark is the minimax
expected length of confidence intervals in $\si_{\gamma, \mathcal
{F}}$,
%
%
\begin{equation}\label{constrained.def}
 L_{\gamma}^*(\mathcal{F}) = \inf_{\mathit{CI} \in
\si
_{\gamma,
\mathcal{F}}}\sup_{f \in\mathcal{F}}E_f(L(\mathit{CI})).
\end{equation}

For convex $\mathcal{F}$, Donoho (\citeyear{Don94}) showed that the modulus of continuity
defined in (\ref{modulus}) determines the minimax expected length,
%
%
\begin{eqnarray}\label{minimax.len}
&&2\omega(2z_{\gamma} n^{-1/2}, \mathcal{F})\nonumber\\ [-8pt]\\ [-8pt]
&&\quad \le
L_{\gamma}^*(\mathcal{F}) \le2 \omega(2z_{\gamma/2}n^{-1/2},
\mathcal{F}),\nonumber
\end{eqnarray}
where
$z_{\gamma}$ is the $100(1 - \gamma)$th percentile of the standard
normal distribution. Moreover, Donoho (\citeyear{Don94}) constructed fixed length
intervals centered at linear estimators which have maximum length
within a small constant factor of the minimax expected length
$L_{\gamma}^*(\mathcal{F})$. Hence, from a minimax point of view
there is
relatively little to gain by centering the intervals on nonlinear
estimators or using variable length intervals.

When the linear functional $T$ is a point evaluation at $t_0\in(0,
1)$, that is, $T(f) = f(t_0)$, and the parameter space is the Besov
ball $\Bb$, the modulus satisfies, with $\nu= \al- {1\over p}$,
\[
\om(n^{-1/2}, \Bb) = C n^{- \nu/(1+ 2\nu)}\bigl(1 + o(1)\bigr).
\]
Following the recipe given in Donoho (\citeyear{Don94}), one can construct a fixed
length $1-\gamma$ level interval centered at a linear estimator with
the length of order $n^{- \nu/(1 + 2\nu)}$.

The situation changes significantly when the parameter space is not
convex. Cai and Low (\citeyear{CaiLow04N1}) developed a minimax theory for parameter
spaces that are finite unions of convex parameter spaces. It is shown
that in this case the optimal (variable length) confidence interval
centered at linear estimators can have expected length much longer
than the minimax expected length; it is thus essential to center the
interval at nonlinear estimators in order to achieve optimality.

When attention is focused on adaptive inference there are some striking
differences between adaptive confidence intervals and adaptive
estimation. As we discussed in the earlier sections, adaptation for
free is often possible under integrated squared error loss and the cost
of adaptation is typically a logarithmic factor under pointwise squared
error loss. For confidence intervals the cost of adaptation can be
substantially more than that for estimation. In fact, in some common
cases, the cost of adaptation is so high that adaptation becomes
basically impossible. In these cases the maximum expected length of the
confidence interval over any parameter space in the collection needs
essentially to be equal to the maximum expected length over the whole
collection in order for the confidence interval to have the desired
coverage probability. See Low (\citeyear{Low97}).

An adaptation theory for confidence intervals was developed in Cai and
Low (\citeyear{CaiLow04N2}). In light of the discussion on adaptive estimation given in
Section \ref{adaptive.sec}, a~natural goal for adaptive confidence
intervals over a~collection of parameter spaces $\{\mathcal{F}_i, i
\in\si
\}$ is to have a~given coverage probability $1-\gamma$ over the union
of the parameter spaces $\mathcal{F}= \bigcup_{i\in\si} \mathcal
{F}_i$ and have the
maximum expected length over each space within a constant factor of the
corresponding minimax expected length, that is,
%
%
\begin{equation}\label{adaptive.EL}
 L(\mathit{CI}, \mathcal{F}_i) \le C_i L_\gamma^*(\mathcal{F}_i),
\end{equation}
where $C_i$ are constants. Unfortunately, in many common cases such
adaptive confidence intervals do not exist even for two parameter
spaces.
Let $\{\mathcal{F}_1, \mathcal{F}_2 \}$ be a~pair of convex parameter
spaces with
nonempty intersection. Let $\mathcal{F}= \mathcal{F}_1 \cup\mathcal
{F}_2$ and $0 < \gamma<
\frac{1}{2}$. It is shown in Cai and Low (\citeyear{CaiLow04N2}) that for $i = 1, 2$
%
%
\begin{eqnarray}\label{EL.bound2}
&&\inf_{\mathit{CI} \in\si_{\gamma,
\mathcal
{F}}}L(\mathit{CI}, \mathcal{F}
_i)\nonumber\\ [-8pt]\\ [-8pt]
&&\quad \ge
\biggl(\frac{1}{2}- \gamma\biggr)\omega_+(z_{\gamma} n^{-1/2}, \mathcal{F}_i,
\mathcal{F}),\nonumber
\end{eqnarray}
where the between class modulus $\omega_+$ is defined\break in~(\ref{b.modulus}).
The lower bound (\ref{EL.bound2}) can in fact be
attained within a~constant factor not depending on $n$. A~general
recipe, which relies on the ordered modu\-lus $\omega(\epsilon,
\mathcal{F}_i,
\mathcal{F}_j)$ defined in (\ref{o.modulus}), is given in Cai and Low (\citeyear{CaiLow04N2})
for the construction of confidence intervals which attains the lower
bound within a constant factor.

The lower bound (\ref{EL.bound2}), however, can be dramatically larger
than the minimax expected length if the parameter space is
prespecified. Such is the case for pointwise confidence intervals over
Besov balls. Consider constructing a confidence interval for a function
at a point $t_0 \in(0, 1)$ over two Besov balls based on the white
noise model. In this case the linear functional $T(f) = f(t_0)$. Let
$\mathcal{F}_i = B^{\al_i}_{p_i,q_i}(M_i)$ with $\nu_i \equiv\al_i
- 1/p_i$
for $i = 1, 2$, $\mathcal{F}=\mathcal{F}_1 \cup\mathcal{F}_2$ and
suppose $\nu_1 > \nu_2>0$.
Then standard calculations, as in, for example, Donoho and Liu (\citeyear{DonLiu}),
show
\begin{eqnarray*}
\omega_+(\ep, \mathcal{F}_i, \mathcal{F}) &=& \omega(\ep, \mathcal
{F})\\
&=& C \ep^{2\nu_2/(1+ 2\nu_2)}\bigl(1+o(1)\bigr),\quad i = 1, 2.
\end{eqnarray*}
Thus, any $1 - \gamma$ level confidence intervals over both
$B^{\al_1}_{p_1,q_1}(M_1)$ and $B^{\al_2}_{p_2,q_2}(M_2)$ must\vspace*{2pt} have the
maximum expected length over $B^{\al_1}_{p_1,q_1}(M_1)$ satisfying
%
%
\begin{eqnarray}\label{len.F1}
&&L(\mathit{CI}, B^{\al_1}_{p_1,q_1}(M_1))\nonumber\\
&&\quad\ge
\bigl(\tfrac{1}{2}- \gamma\bigr)
\om_+(z_{\gamma} n^{-1/2},B^{\al_1}_{p_1,q_1}(M_1),
\mathcal{F})\nonumber\\ [-8pt]\\ [-8pt]
&&\quad\asymp\om(z_{\gamma} n^{-1/2}, \mathcal{F})\nonumber\\
&&\quad\asymp n^{-\nu_2/(1+2\nu_2)}.\nonumber
\end{eqnarray}
In contrast, if it is known that $f \in B^{\al_1}_{p_1,q_1}(M_1)$, $1 -
\gamma$ level confidence intervals\vadjust{\goodbreak} can be constructed which satisfy
\[
L(\mathit{CI}, B^{\al_1}_{p_1,q_1}(M_1)) \le
Cn^{-\nu_1/(1+ 2\nu_1)} \ll Cn^{-\nu_2/(1+ 2\nu_2)}.
\]
From (\ref{len.F1}), the rate of convergence of the maximum expected
length of $\mathit{CI}$ over $B^{\al_1}_{p_1,q_1}(M_1)$ is the same as that for
the maximum expected length over $\mathcal{F}$. From this point of
view the
cost of adaptation is so high that adaptation is impossible.

It is also interesting to note an important difference between
parametric confidence intervals and nonparametric intervals. In the
parametric setting, a universal practice for the construction of a
confidence interval is to first obtain an optimal estimator of a
parameter and then construct a confidence interval for the parameter
centered at this estimator. Such a method often leads to an optimal
confidence interval for the parameter. That is, the confidence interval
has a desired coverage probability and the length of the interval is
the shortest. In nonparametric function estimation, it is also a common
practice to center confidence intervals on optimally adaptive
estimators. However, somewhat surprisingly, this in general leads to
suboptimal confidence procedures (Cai and Low, \citeyear{CaiLow05N3}). That is, either
the confidence interval has poor coverage probability or it is
unnecessarily long. It is instructive to consider an
example.

Let us return to the problem of constructing a~confidence interval for
$f(t_0)$ over the two Besov balls $B^{\al_i}_{p_i,q_i}(M_i)$,\vspace*{2pt} $i=1, 2$.
Again let $\nu_i \equiv\al_i - 1/p_i$ for $i = 1, 2$ and suppose
$\nu_1 > \nu_2> 0$. Equation (\ref{len.F1}) shows that any confidence
interval with coverage probability of at least $1 - \gamma$ over
$\Bbii$ must have the maximum expected length of the order $n^{-\nu_2/(1 + 2\nu_2)}$ over
both $\Bbi$ and $\Bbii$. This bound can easily
be attained by using an optimal fixed length confidence interval. Now
suppose $\hat f(t_0)$ is an adaptive estimator under the mean squared
error. Then, in particular, $\hat f(t_0)$ has the maximum risk over
$\Bbi$ converging at a rate $n^{-r}$ where $r > \frac{2\beta_2}{1 +
2\beta_2}$.\vspace*{2pt} It follows from the results in Cai and Low (\citeyear{CaiLow05N3}) that any
confidence interval $\mathit{CI}$ centered at~$\hat f(t_0)$ with coverage
probability of at least \mbox{$1-\gamma$} over $\Bbii$ must satisfy for some
constant \mbox{$C>0$}
%
%
\begin{eqnarray}\label{EL.bound1}
\hspace*{10pt}L(\mathit{CI}, \Bbii) &\ge& C \biggl(\frac{\log n}{n}\biggr)^{\nu_2/(1 +
2\nu_2)}\nonumber\\ [-8pt]\\ [-8pt]
\hspace*{10pt}&\gg& n^{-\nu_2/(1 + 2\nu_2)}.\nonumber
\end{eqnarray}
Hence, confidence intervals centered at a mean squa\-red error rate
adaptive estimator must have a longer maximum expected length over
$\Bbii$.\vadjust{\goodbreak}

An interesting question is when adaptive confidence intervals exist? It
can be seen easily by comparing the lower bound (\ref{EL.bound2}) with
the bounds (\ref{minimax.len}) for the minimax expected length that
adaptive confidence intervals exist if and only if the moduli satisfy
\[
\omega_+(\ep, \mathcal{F}_i, \mathcal{F}) \asymp\omega(\ep,
\mathcal{F}_i),\quad i =
1, 2,
\]
or, equivalently, $\omega(\ep, \mathcal{F}_2) \le C_1 \omega(\ep,
\mathcal{F}_1)
\le
C_2 \omega_+(\ep,\break \mathcal{F}_1, \mathcal{F}_2)$. In this case
adaptive confidence
intervals exist. These intervals have maximum expected length which can
attain the same optimal rate of convergence as the minimax confidence
interval over\break known~$\mathcal{F}_i$. This is the case for certain shape
restricted function spaces.

Consider constructing pointwise confidence intervals for monotonically
decreasing Lipschitz functions. Again, in this case let $T(f) = f(t_0)$
with $0 < t_0 < 1$. Let\vspace*{1pt} $\mathcal{D}$ be the set of all decreasing functions
on the unit interval and for $0< \beta\le1$ let
%
%
\begin{eqnarray}\label{lipclass}
&&\hspace*{10pt}\mathit{Lip}^\beta(M) = \{f \dvtx[0, 1] \rightarrow\RR,\nonumber\\ [-10pt]\\ [-10pt]
&&\hspace*{10pt}\hphantom{\mathit{Lip}^\beta(M) = \{} |f(x) - f(y)|\le M |x -
y|^{\beta}\}.\nonumber
\end{eqnarray}
Let $\sd^\beta(M) = \sd\cap \mathit{Lip}^\beta(M)$ be the
collection of monotonically decreasing Lipschitz functions. Note that
for $0 < \beta_2 < \beta_1 \le1$, $\sd^{\beta_1}(M) \subset
\sd^{\beta_2}(M)$. Let $\mathcal{F}= \bigcup_{0 \le\beta\le1}\sd
^{\beta}(M)$.
Then standard calculations
yield
%
%
\begin{eqnarray}
&&\hspace*{10pt}\omega_+(\epsilon, \sd^{\beta}(M), \mathcal{F})\nonumber\\[-2pt]
 &&\hspace*{10pt}\quad=\omega(\epsilon,
\sd^{\beta}(M))\\[-2pt]
 &&\hspace*{10pt}\quad= (2\beta+1)^{1/(2\beta+1)}M^{1 /(2\beta
+1)}\epsilon^{2\beta/(2\beta+1)}.\nonumber
\end{eqnarray}
The adaptive confidence interval $\mathit{CI}^*$ given in equation (34) of Cai
and Low (\citeyear{CaiLow04N2}) has coverage probability of at least $1 - \gamma$ over
$\mathcal{F}$ and satisfies for any $0 < \beta\le1$
%
%
\begin{eqnarray}
&&\hspace*{10pt}L(\mathit{CI}^*, \sd^{\beta}(M))\nonumber\\[-2pt]
 &&\hspace*{10pt}\quad\le12 (2\beta+1)^{1/(2\beta+1)}M^{1/(2\beta+1)} z_{\gamma/2}^{2\beta/(2\beta+1)}\\[-2pt]
 &&\hspace*{10pt}\qquad{}\cdot n^{-\beta
/(2\beta+1)}\bigl(1+o(1)\bigr).\nonumber
\end{eqnarray}
Hence, the adaptive confidence interval $\mathit{CI}^*$ simultaneously achieves
with a constant factor of the minimax expected length over all
$\sd^{\beta}(M)$ with $0 < \beta\le1$. Adaptive confidence intervals
also exist for convex functions. See Cai and Low (\citeyear{CaiLow}).

\section{Concluding Remarks}
\label{discussion.sec}

From linear estimators in Pinsker's solution to the ellipsoid problem
to separable rules in Donoho and Johnstone's approach to minimax
estimation over Besov balls to thresholding estimators such as
blockwise James--Stein in adaptive wavelet estimation, shrinkage plays a
pivotal role in both the minimax theory and the adaptation theory in
nonparametric function estimation. In particular, block thresholding
can be viewed as a bridge between the classical normal decision theory
and nonparametric function estimation. Through block thresholding, many
shrinkage estimators developed in the classical theory can be used for
function estimation.

The three problems discussed in the paper are strongly connected. The
minimax difficulty of estimation can be characterized by the modulus
of continuity and the cost of adaptation is captured by the between
class modulus. The linear minimaxity and minimaxity in these three
problems are all linked to the one-dimensional bounded normal mean
problem. In all three problems the performance of linear procedures is
closely linked to the (quadratic) convexity of the parameter space.
Linear shrinkage rules are near optimal when the parameter space is
convex (quadratically convex in the case of global estimation), and
linear procedures can be arbitrarily far from being minimax when the
parameter space is not convex.

Although the minimax theories for the three problems are similar, the
adaptation theories are remarkably different. Among the three problems,
the adaptation results are most positive for estimation under the
global MISE risk. In this case adaptation for free can be achieved. On
the other hand, the results for adaptive confidence intervals are very
pessimistic in general. The cost of adaptation is so high that
adaptation over commonly used smoothness spaces is virtually
impossible, although
adaptation for free can be achieved over shape restricted
spaces. These results indicate that, while the traditional smoothness
constraint works well for estimation, it may not be a practical or
correct formulation for the construction of adaptive nonparametric
confidence intervals or bands. Alternative formulations are needed.
Genovese and Wasserman (\citeyear{GenWas08}) is one step in this direction.

In this paper we have chosen to focus the discussion on the canonical
white noise with drift model to avoid some of the nonessential
technical complications. Parallel results hold for nonparametric
regression and density estimation. We should emphasize that the
discussion as well as the references given in this paper are by no
means extensive. Interested readers are referred to Johnstone (\citeyear{Joh})
for further discussion and for a large number of additional\vadjust{\goodbreak} references
on estimation under global integrated squared error loss.

\section*{Acknowledgments}
We thank the referee for constructive comments which have helped to
improve the presentation of the paper.
Research supported in part by NSF FRG Grant DMS-0854973.\

%

\end{document}